\documentclass[AER]{AEA}
\usepackage{tabularx} 
\usepackage[colorlinks=true, linkcolor=blue, citecolor=blue, urlcolor=blue]{hyperref}
\usepackage{xcolor}
\usepackage{graphicx}
\usepackage{amssymb}
\usepackage{amsmath}
\newtheorem{theorem}{Theorem}[section]
\numberwithin{theorem}{section}
\newtheorem{lem}{Lemma}[section]
\numberwithin{lem}{section}
\newtheorem{ass}{Assumption}[section]
\numberwithin{ass}{section}
\allowdisplaybreaks
\numberwithin{equation}{section}
\usepackage{natbib}
\bibpunct{(}{)}{;}{a}{}{: }
\newcommand{\indep}{\perp \!\!\! \perp}
\usepackage{bbm}

\draftSpacing{1.5}

\begin{document}

\title{Methodological Foundations of Modern Causal Inference in Social Science Research}

\author{Guanghui Pan\footnote{Department of Sociology, University of Oxford. 41 Park End Street, Oxford, the United Kingdom, OX1 1HH. Email: \hyperlink{guanghui.pan@sociology.ox.ac.uk}{guanghui.pan@sociology.ox.ac.uk}. Manuscript in progress, do not cite or redistribute without permission.}}

\maketitle
\section{Notations, and Fundamental Analytical Framework}\label{sec1}

Inference is the core of statistical analysis.  Compared to descriptive analysis of samples, statisticians are more interested in observing a small part of the individuals in a population. Through appropriate inductive reasoning, we can gain knowledge about the population. Therefore, they propose hypotheses and design mathematical models, whether simple or complex, aimed at inferring the characteristics of the population. \\

Mathematically, the population is abstracted as a closed set. Some attributes of the population are \textbf{measurable}. We define \textbf{measurable sets} \(Z\) as a set of non-empty, closed (under complement, countable unions, and countable intersections) subsets of the population; \footnote{which is called \(\sigma-\)algebra. Indeed, suppose the population is \(\mathcal{D}\) and the measurable sets \(Z\), we define a \textbf{measurable space} with the pair \((\mathcal{D}, Z)\). Further, with measure \(\mathcal{P}\), we define the triple \((\mathcal{D}, Z, \mathcal{P})\) as the \textbf{measure space}.}.If we have a set of functions \(\mathcal{P}\) \textbf{measuring} (assign numbers to the characteristics of) the measurable set, or mathematically, mapping the measurable set onto the real numbers (\(\mathcal{P}: Z \to \mathbb{R}\)), we call the functions as \textbf{measures}, or, \textbf{data-generating process (DGP)}.In other words, suppose in a measurable set \(Z\) we have two measurable attributes, \(Z = (X, Y)\); the data-generating process gives out the measurable distributions \(X\) and \(Y\). \\

However, the DGP is, at most times, hard to describe. Consider \(X\) to represent the heights and \(Y\)  to represent the weights of individuals within the British population. It is hard to decipher the data-generating distribution for heights and weights, even if we could possibly obtain them from the census data. An intuitive way to describe it is to set specific \textbf{statistical target parameters} to describe it, for instance, the mean and variance of the height and weight of the British people. The function to map the DGP to the target parameter is called \textbf{estimandor} \(\psi\), and the number of the target parameter \(\psi(\mathcal{P})\) is usually called the \textbf{(statistical) estimand} (i.e., \(\psi(\mathcal{P}) = E_\mathcal{P}(Z)\)).\\

A more common condition is that we could never know the true DGP \(\mathcal{P}\); we could only observe the samples \(Z_i = (X_i, Y_i)\). \textbf{Statistical inference} is the process by which we use the samples to speculate the characteristics of the population. In this thesis, we assume the samples we used in statistical inference are randomly drawn from the population through the same process, or in other words \footnote{Rigorously, the DGP for random samples \(Z_i\) are called, in most textbooks, the random variables. Therefore, the random variable is indeed a function and usually is denoted as \( Z = Z_i\).}: \\
\begin{ass} [Independent and Identically Distributed]\label{ass1}
The samples are \textbf{independent and identically distributed}. \footnote{For instance, using convenience sampling does not draw IID samples. We cannot identify each sample's probability of being sampled from the whole population, and the samples are not connected (consider an extreme scenario in which we sample the heights from a British professional basketball league that can never represent the heights of the British population). Nonetheless, samples from non-IID settings may also perform statistical inference, but I will not address the techniques in this thesis.} \\
\end{ass}

With IID samples, we have the empirical measure (distribution) based on the observations \(\mathbb{P}_n\) (\(n\) denotes the sample size). Correspondingly, we have an \textbf{estimator} denoted as \(\hat{\psi}\) and the empirical \textbf{point estimate} \(\psi(\mathbb{P}_n)\). \footnote{We could understand the estimandor and the estimated separately as the function and the quantity to be estimated, while the estimator and the estimate separately as the function and the quantity we perform estimation based on the observed data. However, in most statistical papers and textbooks, estimator and estimate are interchangeable and rarely rigorously defined, and most researchers assume that the estimator and the estimador have the same functions \(\psi\).  In this thesis, if the notation is clear, I also use \(\psi\) to refer to the estimand \(\psi(\mathcal{P})\) and \(\hat{\psi}\) to refer to the estimator \(\psi(\mathbb{P}_n))\).} If \(\psi(\mathbb{P}_n)\) contains the sample mean, or any linear combination regarding the sample means, under the IID assumption, we have the central limit theorem (CLT):

\begin{theorem}
Central Limit Theorem (CLT): suppose \( \mu\) and \( \sigma^2 \) separately denote the expectation and the variance for the IID observations. If \(\sigma^2\) is finite, as the sample size \(n\) approaches infinity, difference between the average sample mean \(\hat{\psi}\) and the expectation \(\mu\) approaches a normal distribution with mean 0 and variance \(\sigma^2\), at the rate of \(1/\sqrt{n}\):
\begin{align} \label{clt}
    \sqrt{n}\big(\psi(\mathbb{P}_n) - \mu\big) \to \mathcal{N}(0, \sigma^2)
\end{align}
\end{theorem}
If \(\mu\) is an \textbf{unbiased} estimator for \(\psi(\mathcal{P})\), Equation \ref{clt} describes the \textbf{asymptotic} relationship between \(\hat{\psi}\) and \(\psi\), and the difference between  \(\hat{\psi}\) and \(\psi\) is called \textbf{statistical error}. I will elaborate more on the analysis of statistical error in Section \ref{sec3}.\\

Indeed, in some statistical analyses, our ultimate goal is not to estimate the statistical estimand. Instead, we use statistical estimand to approach the "real" estimand in our problems. Consider the following common scenarios in sociological or demographic research: \\
\begin{enumerate}
    \item Suppose we are interested in the \textbf{causal relations} between a treatment and the outcome. Assume that our treatment is binary with only two values: treatment and control. Also, the treatment should be assigned before the outcomes being observed. We use \(Y(1)\) to denote the outcome under treatment, while \(Y(0)\) to denote the outcome under control (in some textbooks, they are denoted as \(Y_t \text{ and } Y_c\)). Let \(A\) denote the treatment variable, and \(X\) denote the set of covariates. The causal measurable set can be written as \(Z^* = (X, A, Y(1), Y(0))\) under the causal DGP \(\mathcal{P}^*\). Suppose our target of interest is the \textbf{average treatment effect} defined as the difference between the expectation of \(Y(1)\) and \(Y(0)\): \(\psi^*(\mathcal{P}^*) = E_{\mathcal{P}^*}\big[E[Y(1)] - E[Y(0)]\big]\). In the real world, due to the fundamental problem in causal inference, we can not observe the outcomes under the treatment and control simultaneously; we could only have the measurable set \(Z = (X, A, Y)\) under the statistical DGP \(\mathcal{P}\) and try to use a reasonable statistical estimand to approach \(\psi^*\), for instance, the conditional expected outcome given the treatment is assigned versus the control is assigned: \(\psi(\mathcal{P}) = E_{\mathcal{P}}\big[E[Y | A = 1] - E[Y | A= 0]\big]\). Only under specific assumptions can we conclude that \(\psi(\mathcal{P})\) is equivalent to \(\psi^*(\mathcal{P^*})\). \\
    
    \item Suppose our parameter of interest is the mean survival time for a (sub)population. A simple measurable data frame is \(Z^* = (X, T)\), in which \(X\) denotes the covariates and \(T\) denotes the survival time (under specific survival probability), and therefore, the estimand is \(\psi^*(\mathcal{P^*}) = E[T]\) \footnote{Of course, in real life analysis, we usually have a more complex data frame like \(Z = (X, S(t_1), S(t_2),..., S(t_n))\), where \(S(t_i)\) denotes the survival probability at \(t_i\). Thus, we first calculate the survival time under specific survival probability (for instance, half-life survival time, in which \(S(t_m) = 0.5)\) with \(S^{-1}\), and average the survival time to get \(E[T]\).}. However, we may encounter the problem that the survival time is censored, meaning that we cannot directly know the survival time \(T\) but know the censoring time \(T_C\). Therefore, the data structure for us is \(Z = (X, T, \delta, T_C)\), where \(\delta\) signals if we occur censoring and \(T_C\) denotes the censoring time. A possible (but uncommon) statistical estimand with the data structure is \(\psi(\mathcal{P}) = E[T | \delta = 0] \), which ignores the censoring data. Further assumptions are required to make our statistical estimand approach the survival estimand. The detailed techniques for survival inference with truncation and censoring are the main part of Section 2 of this thesis. \\
    
    \item Suppose we are interested in how a mediator interferes with the causal relationship between the treatment and the outcome, for instance, how much the causal effect directly goes from the treatment to the outcome (the direct effect) and how much it goes through the mediator onto the outcome (the indirect effect) \citep[Chapter~2]{vanderweele2015}. We still assume that our treatment is a binary one \(A = 1\) denotes the treatment, and \(A=0\) denotes the control. The mediator takes the value of \(m(1)\) when \(A = 1\) and \(m(0)\) when \(A = 0\). Thus, there are four combinations of the potential outcome: \(Y(1, m(1)), Y(1, m(0)), Y(0, m(1)),\) and \(Y(0, m(0))\). Our estimands are the two components of the total average treatment effect \(E[Y(1,m(1))] - E[Y(0,m(0))]\): the direct effect, defined as the average effect of treatment in the absence of the mediator \(E[Y(1, m(0))] - E [Y(0, m(0))]\); and the indirect effect, defined as the difference between the causal effects with and without the mediator \(E[Y(1, m(1))] - E [Y(1, m(0))]\). Indeed, the most important parts, as shown in the decomposition, are the \textbf{conditional response function}: \(E[Y(a,m)]\) and we let the target mediation estimator \(\psi^*(\mathcal{P}^*)\) to be that. From our statistical model, in which we have the data formed in a tuple: \(Z = (X, A, M, Y)\), we could yield the conditional expectation term, \(\psi(\mathcal{P}) = E[Y|A = a, M = m]\). Similarly, we need specific assumptions that allow us to equalize \(\psi^*(\mathcal{P}^*)\) and \(\psi(\mathcal{P}) \). The details of the technique are discussed in Section 3 of this thesis. \\
\end{enumerate}

Since the content of causal inference (the first scenario above) goes throughout the thesis, we specifically analyze it in this chapter. \textbf{Causal inference}, broadly speaking, is a process that determines an independent effect of a particular object (the treatment) on another (the outcome), and it is usually contained in a larger system. For instance, when Galileo experimented on the Leaning Tower of Pisa, he isolated the independent effect of the weights of the two balls, observing that the heavier ball and the lighter ball fell on the ground simultaneously, and therefore inferred that the weights (masses) of the balls (the treatment) have nothing to do with the gravity (the outcome) of the two balls. The strategy to isolate the treatment effect is a\textbf{ controlled experiment}, in which we are assured that the only difference between the two is the object we deem the treatment. However, in many scientific disciplines, we can not artificially manipulate and completely isolate the treatment\footnote{Indeed, in Galileo's experiment, the masses could not be the only difference between the two balls, as either the materials (densities) or the size must be different because the densities and the volumes determine the masses (in Galileo's original experiment he ensured the two balls were irony). Therefore, further experiment designs are needed: controlling the size of the two balls and controlling the type of materials of the two balls.}. Moreover, due to the requirement of repeatability in modern science, we usually observe the \textit{group-level} differences between treated and untreated instead of the two individuals (balls). Based on this, experiment designers need techniques such as \textbf{randomization} or \textbf{blind control} for treatment assignment \footnote{Randomization and blind-control might refer to different techniques. Consider the example of Pavlov's classic conditioning experiment with the dog, which is a blind instead of randomized control.}. Experiments with randomized group assignment are commonly called \textbf{Randomized Control Trials (RCT)}.

Experiments are becoming increasingly common in social science studies, but most research still relies on \textbf{observational data} to infer causal relationships. The observational data does not have the RCT design: they do not randomize the samples to the treatment and control groups. They are not designed to isolate the effects of the treatment variable. Therefore, researchers need to use statistical techniques to transfer the observational cases to approximate the RCT process, and therefore, causal inference with observational data is a \textbf{pseudo-RCT}. Our thesis mainly addresses statistical ideas on causal inference with observational data. 

In this thesis, causal inference with observational data is the process using the observable estimator from the random samples \(\psi(\mathbb{P}_n)\) to estimate the causal estimand \(\psi^*(\mathcal{P^*})\) via the statistical estimand \(\psi(\mathcal{P})\). In empirical studies, the estimation functions for the estimand (population distribution) (estimandor) and for the observables (sample distribution) are always the same (for instance, the expectation  \(\psi(\mathcal{P}) = E_{\mathcal{P}}[Z]\) and \(\hat{\psi}(\mathbb{P}_n) = E_{\mathbb{P}_n}[Z_i]\) have the same functional form), and the only difference between the two functions here is in the measurement choice. Thus, below, we will use \(\hat{\psi}\) and \(\psi({\mathbb{P}}_n)\) interchangeably to represent the estimator from the empirical dataset, we use \(\psi\) and \(\psi(\mathcal{P})\) interchangeably to represent the estimator (estimandor) from the statistical estimand.

Statisticians prefer \(\hat{\psi}\) as an unbiased estimator on \(\psi^*\) and regard the divergence between the two terms as error terms. Moreover, we can decompose the error term into the statistical error, which is the divergence between the estimator and the statistical estimand, and the causal error, which is the divergence between the statistical estimand and the causal estimand:
\begin{align} \label{totalerror}
    \hat{\psi} - \psi^* = \underbrace{(\hat{\psi} - \psi)}_{\text{statistical error}} + \underbrace{(\psi - \psi^*)}_{\text{causal error}}
\end{align}
For the statistical error, we could further decompose it into the statistical variance and the statistical bias:
\begin{align}\label{staterror}
    \hat{\psi} - \psi = \underbrace{(\hat{\psi}- E[\hat{\psi}])}_{\text{statistical variance}} + \underbrace{ (E[\hat{\psi}] - \psi)}_{\text{statistical bias}}
\end{align}
Analyzing the error terms is the core of the causal analysis, as in substantive work, especially from observational data, our estimator at most times could not be satisfied as unbiased yielding the causal estimand (and even in the RCT, further assumptions may be required). In Section \ref{sec2}, we discuss the causal error, and in Section \ref{sec3}, we discuss the statistical error. 

\section{Analyzing the Causal Error}\label{sec2}
\subsection{Unbiased estimators and assumptions}

The causal error is defined as the divergence between the statistical estimand \(\psi(\mathcal{P})\) (simply \(\psi\)) and the causal estimand \(\psi^*(\mathcal{P^*})\) (simply \(\psi^*\)). We cannot directly apply the statistical estimand as the causal one because of the \textbf{fundamental problems of causal inference} that we do not observe the outcome under different treatment conditions simultaneously, as each individual only receives one identifiable treatment. In \textbf{Neyman-Rubin's (NR)} causal framework \citep{neyman1935, rubin1990}, the outcomes under different treatment statuses from the causal DGP (\(Y(a))\) are the \textbf{potential outcomes} \citep{holland1986}. In the statistical data frame, the existing outcome is conditioned on the assigned value of the treatment (\(Y | A= a\)); nevertheless, we could not directly obtain the outcomes conditioned on the treatment assigned to other values. Therefore, the unobservable outcomes are called \textbf{counterfactuals}. \\

NR's counterfactual framework is not the only way to understand the causal and statistical estimand relationship. Computer scientist Judea Pearl (\citeyear{pearl2009}) created a framework called \textbf{"do-calculus" (DoC)}. From Pearl's perspective, the outcome is deterministic if the treatment action has been triggered and the mapping rule from the treatment to the outcome is determined: if the treatment is \(A = a,\), then \(a \to Y(a)\); if \(A = a^{'}\), then \(a^{'} \to Y(a^{'})\). Based on the mapping symbol, Pearl developed the graphical expression for causal analysis called the \textbf{Direct Acyclic Graphs (DAG)}, and it is commonly used in structural equation models and causal mediation analysis. \\

Indeed, both NR's counterfactual and Pearl's DoC framework require additional assumptions to make \(\psi_a^*(\mathcal{P}^*) = E_{\mathcal{P}^*}[Y(a)]\) and \(\psi_a(\mathcal{P}) = E_P[Y | A = a]\) equivalent. In this whole thesis, we consider the binary treatment. We suppose that the probability of being assigned as treatment and control is a positive number between 0 and 1:
\begin{ass}[Positivity Assumption]\label{positivity}
    \footnote{In some literature, this assumption is also called overlap assumption \citep{heckman1998}.} The probability of being assigned as treatment and control is a positive number between 0 and 1:
    \[
    P(A = 1) \in (0,1); P(A = 0) \in (0,1)
    \]
    Where \(A\) is the treatment. 
\end{ass}

And suppose that in the "omniscient" causal data frame \(Z^* = (X, A, Y(1), Y(0)\), we could observe the following conditional expectations: \(E[Y(1) | A = 1]\), \(E[Y(1) | A = 0]\),\(E[Y(0) | A = 1]\),  \(E[Y(0) | A = 0]\), while in the statistical data frame \(Z = (X, A, Y)\) we could only observe \(E[Y | A = 1]\) and \(E[Y | A = 0]\). With the following assumption, we could link between \(E[Y | A = 1]\) and \(E[Y(1) | A = 1]\), and between \(E[Y | A = 0]\) and  and \(E[Y(0) | A = 0]\):
\begin{ass}[Consistent Assumption]\label{consistent}
    The potential outcome under treatment received is the same as the observed outcome. That is,
    \[
    Y = Y(A)
    \]
    where \(Y\) is the observed outcome, \(Y(A)\) is the potential outcome, and \(A\) is the treatment.
\end{ass} 

Indeed, in the DoC framework, consistency has been implied since the mapping function from the execution of the treatment to the outcome is defined. Also, it is worth noting that when we use observational data to infer the causal estimand, consistent assumption needs to be held on the individual level: \(Y_i = Y_i(A_i)\). Furthermore, since we have Assumption \ref{ass1} for the IID samples, we could infer that there's no interference among individuals: the treatment assigned to one observational sample does not affect the outcome of the others. Consistency and no interference assumptions on the individual observational level are collectively known as the \textbf{Stable Treatment Unit Value Assumption}, or SUTVA. \\

If we have established the relationship between the causal and statistical data frames, that \(E_\mathcal{P}[Y | A = a]\) = \(E_{\mathcal{P}^*} [Y (a) | A = a]\). We suppose the proportion assigned to the treatment group is \(\rho\), and therefore, \(E[Y(1)] = \rho E[Y(1) | A = 1] + (1-\rho)E[Y(1) | A = 0]\) and \(E[Y(0)] = \rho E[Y(0) | A = 1] + (1-\rho)E[Y(0) | A = 0]\)
We may calculate the difference between the causal average treatment effect and the statistical treatment effect with a simple calculation: 
\begin{equation} \label{errordecompose}
\begin{aligned}
    &\underbrace{\big(E[Y(1) | A = 1] - E[Y(0) | A = 0]\big)}_{\text{statistical average treatment effect}} - \underbrace{\big(E[Y(1)] - E[Y(0)]\big)}_{\text{causal average treatment effect}}  \\
    &= (E[Y(1) | A = 1] - E[Y(0) | A = 0]) \\
    &\quad -\bigg[\big(\rho E[Y(1) | A = 1] + (1-\rho)E[Y(1) | A = 0]\big) \\
    &\quad -\big(\rho E[Y(0) | A = 1] + (1-\rho)E[Y(0) | A = 0]\big)\bigg] \\
    &= \underbrace{E[Y(0) |A = 1] - E[Y(0) | A = 1]}_{\text{difference in baseline}} + \underbrace{(1-\rho)(\delta_1 - \delta_0)}_{\text{heterogeneous treatment effect}}
\end{aligned}
\end{equation}
\noindent where
\begin{align*}
    \delta_1 = E[Y(1) | A = 1] - E[Y(0) | A = 1] \quad \text{and} \quad \delta_0 = E[Y(1) | A = 0] - E[Y(0) | A = 0]. 
\end{align*}
\begin{proof}
Let \(E[Y(1) | A = 1] = \alpha_1\), \(E[Y(1) | A = 0] = \alpha_2\), \(E[Y(0) | A = 1] = \alpha_3\), and \(E[Y(0) | A = 0] = \alpha_4\). 
    Therefore, the left side of the equation is:
    \[
    (\alpha_1 - \alpha_4) - \left[\rho\alpha_1 + (1-\rho)\alpha_2 - \rho\alpha_3 - (1-\rho)\alpha_4\right].
    \]
    Simplifying this, we have:
    \begin{align*}
        (\alpha_1 - \alpha_4) 
        &- \left[\rho\alpha_1 + (1-\rho)\alpha_2 - \rho\alpha_3 - (1-\rho)\alpha_4\right] \\
        &= (\alpha_1 - \alpha_4) - \rho\alpha_1 - (1-\rho)\alpha_2 + \rho\alpha_3 + (1-\rho)\alpha_4 \\
        &= \alpha_1 - \alpha_4 - \rho\alpha_1 - (1-\rho)\alpha_2 + \rho\alpha_3 + (1-\rho)\alpha_4 \\
        &= \underbrace{(\alpha_3 - \alpha_4)}_{\text{difference in baseline}} + (1-\rho)\underbrace{(\alpha_1 - \alpha_3) - (\alpha_2 - \alpha_4)}_{\text{heterogeneous treatment effect}}.
    \end{align*}
    Rewriting this back in terms of the original expectations, we have the right side of the equation.
\end{proof} 

Equation \ref{errordecompose} reveals the two origins of bias in causal inference when using the statistical estimator to infer the causal estimand: the baseline difference, or the selection bias, which is the pre-treatment divergence when grouping individuals to the treatment and control groups; and the heterogeneous treatment effect between the treatment and control group, which is the post-treatment divergence between the treatment and the control group.  For example, consider evaluating the impact of a training program on workers' productivity. Initially, we measure their productivity levels before the training. Next, we divide the workers into two groups: a treatment group that receives the training and a control group that does not. After a certain period, we measure the change in productivity in both groups to assess the effect of the training program. The bias in this measurement comes from two sources: firstly, a pre-training bias, where workers in the treatment group might have different initial productivity levels compared to those in the control group; and secondly, a post-training bias, where workers in the treatment group might experience a greater improvement in productivity than those in the control group, even if they had all received the training.\\

Therefore, to eliminate the potential pre and post-treatment bias, we need further assumptions for identification. Since the pre-treatment selection and post-treatment heterogeneity can be attributed to the non-randomization in the treatment assignment, we have the ignorability/ unconfoundedness assumption: 
\begin{ass}[Ignorability/Unconfoundedness Assumption]\label{unconfound}
    The treatment assignment \(A\) is independent to the potential outcomes Y(1) and Y(0):
    \[
    Y(1), Y(0) \indep A \footnote{The symbol \(\indep\) is the independent symbol, it means the two objects (vectors or matrices) are uncorrelated: \(a \indep b \iff cov(a,b) = 0 \iff E[a \mid b] = 0\). The three expressions are exchangeable in our use.}
    \]
\end{ass}
Since the potential outcome is independent of the assignment of the treatment, we can infer that \(E[Y(1) | A = 1] = E[Y(1) | A = 0] \) and \(E[Y(0) | A = 1] = E[Y(0) | A = 0] \). Therefore, under the unconfoundedness assumption, the pre-treatment baseline difference is \(0\). Meanwhile, the gap between \(E[Y(1) | A = 1]\) and \( E[Y(0) | A = 1] \) is the same as the gap between \(E[Y(0) | A = 1]\) and \( E[Y(0) | A = 0] \), eliminating the post-treatment heterogeneity\footnote{The elimination of post-treatment heterogeneity with the unconfoundedness assumption does not eliminate what econometrists called the heterogeneous treatment effect (HTE) in causal inference. First, although the average post-treatment heterogeneity between the treatment and control groups is eliminated, the individual treatment effect \(Y_i(0) |A_i = 1\) and \(Y_i(1) | A_i = 0\) still exists. Secondly, the HTE is indeed a conditional expectation \(E[Y(1) | X] - E[Y(0) |X]\) instead of what the unconfoundedness assumption controls \(E[Y(1)] - E[Y(0)]\).}. In this sense, with Assumptions \ref{positivity},\ref{consistent}, and \ref{unconfound}, we could finally conclude that the statistical average treatment effect is an unbiased estimand on the causal average treatment effect: \(\psi(\mathcal{P}) = \psi^*(\mathcal{P}^*)\).  \\

In most circumstances, indeed, we may find a set of covariates \(X\) in the statistical model are correlated with both \(A\) and \(Y\), violating the ignorability/unconfoundedness assumption. Thus, we may randomize the treatment assignment conditioned on \(X\), which is a \textbf{pseudo-randomization}. We update Assumptions \ref{positivity} and \ref{unconfound} to make them include the conditions of the covariates (the consistency hypothesis remains unchanged): 
\begin{ass}[Causal Inference Assumptions]\label{ciass}
    Suppose a statistical DGP \(Z = (X, A, Y)\), in which \(X\) denotes the covariates, \(A\) denotes the treatment, and \(Y\) denotes the outcome. To make the statistical estimand \( \psi(\mathcal{P}) = E[E_X[Y | A = 1, X]] - E[E_X[Y | A = 1, X]]\) equivalent to the causal estimand \( \psi^*(\mathcal{P}^*) = E[Y(1)] - E[Y(0)] \) from the causal DGP \(Z^* = (X, A, Y(1), Y(0))\) (where \(Y(1)\), \(Y(0)\) denote the potential outcomes under treatment and control, respectively), we need the following hypotheses: 
    \begin{enumerate}
        \item Positivity: the probability to be assigned to treatment and control group conditioned on the covariates, is a positive number between 0 and 1:
        \[
        P(A = 1 |X) \in (0,1); P(A = 0 | X) \in (0,1)
        \]
        \item Consistency: the potential outcome under the treatment received is the same as the observed outcome: 
        \[
        Y = Y(A)
        \]
        \item Unconfoundedness: conditional on a set of observed covariates \(X\), the potential outcomes \(Y(1)\) and \(Y(0)\) are independent of the treatment assignment \(A\):
        \[
        \{ Y(1), Y(0) \} \indep A | X
        \]
    \end{enumerate}
\end{ass}
Assumption \ref{ciass} is the sufficient and necessary condition for the statistical estimand on the average treatment effect to be equivalent to the causal estimand on the average treatment effect. To simplify, consider our target parameter is the potential outcome \(\psi_a^* = E[Y(a)]\), and our statistical estimand is \(\psi_a = E[E_X[Y | A = a, X]]\) (as \(E_X[Y(a) |X ] = \int y f_{Y(a) |X}(y |X)dy\)), we have:
\begin{equation}\label{expectation}
    \begin{aligned}
        \psi_a^* &= E[Y(a)] \\
        &= E[E_X[Y(a) |X]] \text{ (conditional expectation)} \\
        &= E[E_X[Y(a) | A, X]] \text{ (positivity and unconfoundedness)} \\
        &= E[E_X[Y | A= a, X]] \text{ (consistency)} \\
        &= \psi_a 
    \end{aligned}
\end{equation}
Meanwhile, Equation \ref{expectation} can be also written as: 
\begin{equation} \label{weights}
    \begin{aligned}
    \psi_a^* &= E[Y(a)] \\
        &= E[E[Y(a) |X]] \text{ (conditional expectation)} \\
        &= E\bigg[E[Y(a)|X] \frac{E[\mathbbm{1}_A |X]}{E[\mathbbm{1}_A |X]}\bigg] \text{ (positivity;} \mathbbm{1}_A = 1 \text{ if } A = a; 0 \text{ otherwise)} \\
        &= E\bigg[\frac{E[Y(a) \mathbbm{1}_A \mid X]}{E[\mathbbm{1}_A |X]}\bigg] \text{ (unconfoundedness)} \\
        &= E\bigg[\frac{E[Y \mathbbm{1}_A \mid X]}{P[A = a\mid X]}\bigg] \text{ (consistency)} \\
        &= E\bigg[\frac{E[Y \mathbbm{1}_A \mid X]}{\pi_a(X)}\bigg] \text{ (define } \pi_a(X) = P(A= a|X)\text{)} \\
        & = E\bigg[\frac{Y \mathbbm{1}_A}{\pi_a(X)}\bigg] \text{ (reverse conditional expectation)}
    \end{aligned}
\end{equation}
Equations \ref{expectation} and \ref{weights} illustrate that we could infer the potential outcome \(\psi_a^*\) with either the conditional expectation \(\psi_a\) or the propensity score function for the treatment \(\pi(a)\) as the unbiased estimators. The results are the foundation of what we refer to as the \textbf{doubly robust/debiased estimation} later in this chapter.

\subsection{Violations on causal assumptions}
In most social science research scenarios with observational (survey) data, finding an unbiased causal estimator is challenging, as the three conditions in Assumption \ref{ciass} are not always satisfied, especially the positivity and the unconfoundedness assumptions. 
\subsubsection{Violations on positivity}
In social science, violating the positivity assumption is common if our treatment involves policy/reform/law enforcement that affects all our research objects. For instance, suppose our target is to measure how the rules on sports gambling may affect suicide risks for the residents of a state. Since the law affects everyone in the state, it violates the positivity assumption as \(P(A) = 1\) (everyone in the state is grouped as treated). Violating the positivity assumption will make the casual estimand \textbf{unidentifiable}, as potential outcomes under counterfactual scenarios are nonexistent. Therefore, in the situations discussed above, further assumptions are required for causal inference.
\begin{itemize}
    \item \textit{Difference in Differences} \footnote{It is worth noting that methods mentioned here may be applied to scenarios which violate the other assumptions, or under which no assumption is violated.} \\
    A common technique to address the causal inference if the positivity assumption is violated is the \textbf{difference in difference (DID)} method (especially with longitudinal data), with additional assumptions and control group settings. For instance, in the example above, we may artificially choose the comparable control group as a neighboring state with similar socioeconomic factors but without changes to sports gambling laws. In this sense, we have implicitly assumed that the consistency and unconfoundedness assumptions to be true, since the treatment state and the control state will not affect the outcome of each other, and there are no other covariates affecting the change in suicide rates in both states during the observational time. Additionally, we need to assume that the suicide rate trend in the control state does not change before and after the change of the gambling law. In other words, it is called the \textbf{parallel assumption} \citep{roth2023}:
    \begin{ass}[ Parallel Assumption for Difference in Differences]\label{parallel}
       In the absence of the treatment, the difference in the average outcomes between the treated and control groups remains constant over time:
        \[
        E[Y_{t_1}(0) \mid A = 1] - E[Y_{t_0}(0) \mid A = 1] = E[Y_{t_1}(0) \mid A = 0] - E[Y_{t_0}(0) \mid A = 0]
        \]
        where \(Y_{t_1}\) and \(Y_{t_0}\) separately denotes the outcome before and after the treatment. 
    \end{ass}
    With Assumption \ref{parallel}, we indeed transfer the causal effect of the policy change into the causal effect of time. With the introduction of a "comparable" control group, the probability of receiving treatment becomes a real number between 0 and 1. Meanwhile, since \(E[Y_{t_1}(0) | A = 1]\) is unobservable from the observational data, the assumption is not testifiable. The causal estimand, in this regard, should be the difference between the observable expected post-treatment outcome and the counterfactual post-treatment outcome (assuming the treatment was not received):
    \[
    E[Y_{t_1}(1) | A = 1] - E[Y_{t_1}(0) | A = 1] 
    \]
    Although the assumption is not testifiable, we can still justify the effectiveness of using the DID method. For instance, we can use statistical or visualization methods for the \textbf{palacbo test}, which examines whether the trends for the treatment group and the control group before the treatment were paralleled. Moreover, we could also apply a \textbf{falsification test}, which sets an outcome that is impossible to be affected by the treatment, to see if the trend for the outcome changes in the treatment group.  \\

    DID framework can be extended to the \textbf{Difference in difference in differences (DDD)}. In the DID framework, we assume that the unconfoundedness assumption is not violated, as we adjust for the time-invariant differences between groups and common trends affecting all units. Meanwhile, the unconfoundedness assumption may not hold, as some confounding variables might vary across groups and over time and further affect the parallel assumption. For instance, assuming we would like to evaluate the effectiveness of a specific educational policy on students' academic performance and choose a province for the experiment to further promote it to the whole nation, our treatment effect should be inferrable. Suppose the experimental province has only implemented it in its cities but not in the rural areas; if we only use the DID method to compare the effectiveness with the neighbor province while not noticing that we mixed the heterogeneous effects between urban and rural areas, our estimation may somehow underestimate the effectiveness. Therefore, we need to differentiate the DID results on the layer of urban or rural areas and approximate the causal effects more precisely. 
    
    \item \textit{Regression Discontinuity} \\
Like DID, \textbf{Regression Discontinuity (RD)} is another method that, with additional assumptions of continuity at the breakpoint, addresses violations of positivity by assuming that the probability of treatment after the breakpoint is not always 1. For instance, suppose our target is to evaluate how tax cuts stimulate consumption for those with an annual income exceeding \$100,000. Imagine the fiscal policy increases the tax rate for individuals earning beyond \$100,000 from 25\% to 30\% (while remaining constant for people earning below \$100,000 at 25\%). Without further assumptions, we cannot identify the policy's effect because all individuals earning beyond \$100,000 are assigned to the treatment group, violating the positivity assumption. As demonstrated in the DID section, we could assume the parallel trend to compare the treated and untreated groups (who earn below \$100,000) before and after the treatment and identify the causal effect of the tax reform on consumption. \\

Moreover, we could have a continuity assumption under the RD setting, assuming the distribution is continuity at the policy's breakpoint. In the example above, we assume that there's a slight (no) difference between the consumption behaviors for those earning \textit{just} below and \textit{just} above \$100,000 (how much the \textit{just} is the radius of the threshold). 
\begin{ass}[Continuity Assumption for Regression Discontinuity]\label{continuity}
     
The expected potential outcomes \( E[Y(0) | X = x] \) and \( E[Y(1) | X = x] \) are continuous at the threshold \( c \):
\[ \lim_{x \to c^-} E[Y(0) | X = x] = \lim_{x \to c^+} E[Y(0) | X = x] \]
\[ \lim_{x \to c^-} E[Y(1) | X = x] = \lim_{x \to c^+} E[Y(1) | X = x] \]
Where the potential outcomes are denoted as \( Y(1) \) if treated, and \( Y(0) \) if untreated. \(X_i\) denotes the running variable.
\end{ass} 
Based on Assumption \ref{continuity}, we made up the potential outcome where \(0 < \lim_{x \to c^-}P(Y(1) |X = x) < 1\) and \(0 < \lim_{x \to c^+}P(Y(1) |X = x) < 1\) to get over the positivity assumption. Therefore, due to Assumption \ref{ciass}, we could assign the statistical estimand below the threshold as the causal estimand for the untreated group, whereas the statistical estimand above the threshold as the causal estimand for the treatment group:
\[ \lim_{x \to c^-} E[Y | X = x] = \lim_{x \to c^-} E[Y(0) | X = x] \]
\[ \lim_{x \to c^+} E[Y | X = x] = \lim_{x \to c^+} E[Y(1) | X = x] \]
and the difference in the conditional expectations of the observed outcomes on either side of the cutoff can be attributed to the treatment effect:
\[ \lim_{x \to c^+} E[Y | X = x] - \lim_{x \to c^-} E[Y | X = x ]=\lim_{x \to c^+} E[Y(1) | X = x] - \lim_{x \to c^-} E[Y(0) | X = x] \]. \\
In the example above, suppose we take a radius of \$5,000 and the threshold is \$100,000. Therefore, we may calculate the average consumption costs for people after the tax reform earnings between \$95,000 to \$100,000 and for those earnings between \$100,000 to \$105,000. The treatment effect can be captured by the difference between the two expected consumption costs. \\

Applying the RD method has some restrictions on the running variable \(X\) and potential confounding variables. For the running variable, we have to assume that the patterns in the radius of the threshold are consistent\footnote{In some literature, this is also called "no manipulation" to ensure the continuity below and above the cutoff point.}. In the above example, if the tax rates originally were different for people earning below and above \$100,000 (for instance, originally, people earning below \$100,000 only had a tax rate of 15\% and remained constant after the reform), the RD method requires further auxiliaries for identification. Or if the original tax rate cutoff point was at \$97,500 (people earning below \$97,500 received 15\% tax rate while those earning above \$97,500 received 25\% tax rate), we need to redesign the bandwidth of the radius around the threshold. For the covariates that may interfere with the potential outcomes, the RD method also requires them to cross the threshold smoothly. In this way, the only difference between people under and above the threshold is whether they are assigned to the treatment group (above the threshold). In other words, the continuity assumption does not ensure the unconfoundedness condition is satisfied: as our estimand \(\lim_{x \to c^+} E[Y(1) | X = x] - \lim_{x \to c^-} E[Y(0) | X = x]\) is indeed the statistical estimand \(E[Y(1) | A = 1] - E[Y(0) | A = 0] \) in Equation \ref{errordecompose} and we still have the bias into baseline difference and heterogeneous treatment effect. For instance, in the previous example, we need to ensure that educational levels for people earning between \$95,000 to \$100,000 and \$100,000 to \$100,500 are almost the same, excluding the interference from the covariate of educational levels. 
\end{itemize}
The positivity assumption requires that the probability assigned to the treatment and control group is a real number between 0 and 1, which, if violated, will make the causal estimand unidentifiable. DID and RD are, with additional assumptions, appropriate ways to construct the positivity in the probability of assigning cases to the treatment and control groups. Both methods are also helpful when addressing the violation of the unconfoundedness assumption-- ignoring the possible interference from the unequal distribution of the covariates between the treatment and the control groups. 
\subsubsection{Violations on unconfoundedness}

Unlike the violation of the positivity assumption, which makes the causal effect unidentifiable, the violation of the unconfoundedness may only yield bias for estimating the causal effect. As Equation \ref{errordecompose} suggests, when the unconfoundedness assumption does not hold, the bias using statistical estimand to infer the causal estimand can be attributed to the pre-treatment selection bias and the post-treatment heterogeneity. The unconfoundedness assumption states that the outcome \((Y(1), Y(0))\) is independent of the treatment \(A\) given a set of covariates \(X\).In econometrics terms, we call treatment \(A\) the endogenous treatment, while covariates \(X\) are exogenous covariates. Based on the relationship among \(X, Y, \text{and} A\), previous researchers developed two intrinsically consistent ways to address the violation of the unconfoundedness assumption: the propensity score function-based matching and weighting method and eliminating exogeneity based on the local treatment average effect method, with instrumental variables and fixed effects. 
\begin{itemize}
    \item \textit{Propensity Function based Matching and Weighting}\\
    
With covariates affecting group assignment known, matching or weighting is the intuitive choice to eliminate the unconfoundedness bias. If covariates \(X\) are all known, as Equation \ref{unconfound} indicates, with the intermediate estimator \(\pi_a(X) = P(A = a \mid X)\), we can have the unbiased inference of the causal estimand \(E[Y(a)]\). Traditionally, we call the intermediate estimator \(\pi_a(X)\) as the \textbf{propensity score} function, as it measures the likelihood of treatment group assignment given the covariates \(X\). The expression using the identity function divided by the propensity score function \(\frac{\mathbbm{1}_a}{\pi_a(X)}\) is called the \textbf{inverse probability weighting (IPW)}. Therefore, the IPW method yields the unbiased (and also regular and asymptotically linear, see Section \ref{sec3}) estimation for causality if all assumptions hold. \\

The IPW method achieves pseudo-randomization since the reweighting process reassigns individuals into the pseudo-treatment groups based on their propensity (likelihood) rather than their true assignment status (which lacks the randomization process). The precondition for weighting is that covariates \(X\) affect the propensity and cause selection bias if not controlled. Therefore, if we have a longitudinal study, \(X\) should be variables \textit{ex-ante} the treatment assignment. Suppose we have a set of variables \(M\) that affect the outcome \textit{ex-post} of the treatment assignment. In that case, they will not affect our estimation of the causal effect \(A \to Y\). Still, they will decompose the total treatment effect into the direct one (the treatment directly affects the outcome) and the indirect (the causal effect goes via \(M\) onto \(Y\)) effects, which we will discuss in the mediation section\footnote{The unconfoundedness assumption states \((Y(1), Y(0)) \indep A \mid X \iff cov(Y, A\mid X) = 0\), which does not indicate either \(cov(Y, A) = 0\), \(cov(X, A) = 0\), or \(cov(Y, X) = 0\). Instead, \(cov(Y, A) \ne 0\) and \(cov(X, A) \ne 0\) are the foundations of causal inference (especially for Equation \ref{unconfound}. The relationship between \(Y\) and \(X\) is the trickiest: \(cov(Y, X)\) is not necessary \(0\), but  \(cov(Y, A \mid X) \equiv 0\) states that there might be some correlations between \(Y\) and \(X)\), but such correlation has to be blocked under treatment \(A\). Under this circumstance, the treatment assignment is randomized given \(X\). In the next section, we will see that \(X\) is indeed the instrumental variable.}. \\

Based on the propensity score function, weighting with inverse probability (propensity) provides a way to manipulate the treatment assignment to attain pseudo-randomization. Similarly, a matching method based on the propensity score function \footnote{Propensity score function is not a score function that will be introduced in Section \ref{sec3}. Therefore, to avoid confusion, in the following part of this chapter, we only call it "propensity function."} can also achieve the effect of pseudo-randomization. This is the classic \textbf{propensity score matching (PSM)} method for causal inference \citep{rosenbaum1983}. Suppose we could match the cases assigned in the treatment group and the control group with exactly the same propensity score and calculate the difference between the matched cases along the propensity score spectrum, and we finally calculate the average; we could yield the causal estimand for the average treatment effect based on the matching method. \\

The method seems plausible, but when dealing with real observational data, researchers have to make a tradeoff between the quality of the matching algorithm and the selection of cases. The real problem is that we rely on the observational data to generate the estimator \(\hat{\pi}_a\) to estimate \(\pi_a\).  We can imagine that when using the observational data to estimate the propensity function, for the individuals in the treatment group, the distribution is likely to be dense at the end towards \(1\) (if \(1\) indicates being assigned to the treatment group) and relatively sparse at the end towards \(0\), while for the control group individuals tend to distribute denser on the side of \(0\) and more sparse on the side of \(1\). Therefore, it is infeasible to have a one-on-one match between the individuals from the treatment group and the control group, with the exact same propensity value, and get everyone matched (see Figure \ref{fig:psm} for the illustration). Researchers have to adopt methods either to allow the divergence (caliper) in propensity scores between the matched cases, to drop the unmatched cases, or a method with the combination of the two (for instance, set a threshold for nearest neighborhood matching and drop the cases beyond the threshold).\\

\begin{figure}[ht]
    \centering
    \includegraphics[width=\textwidth]{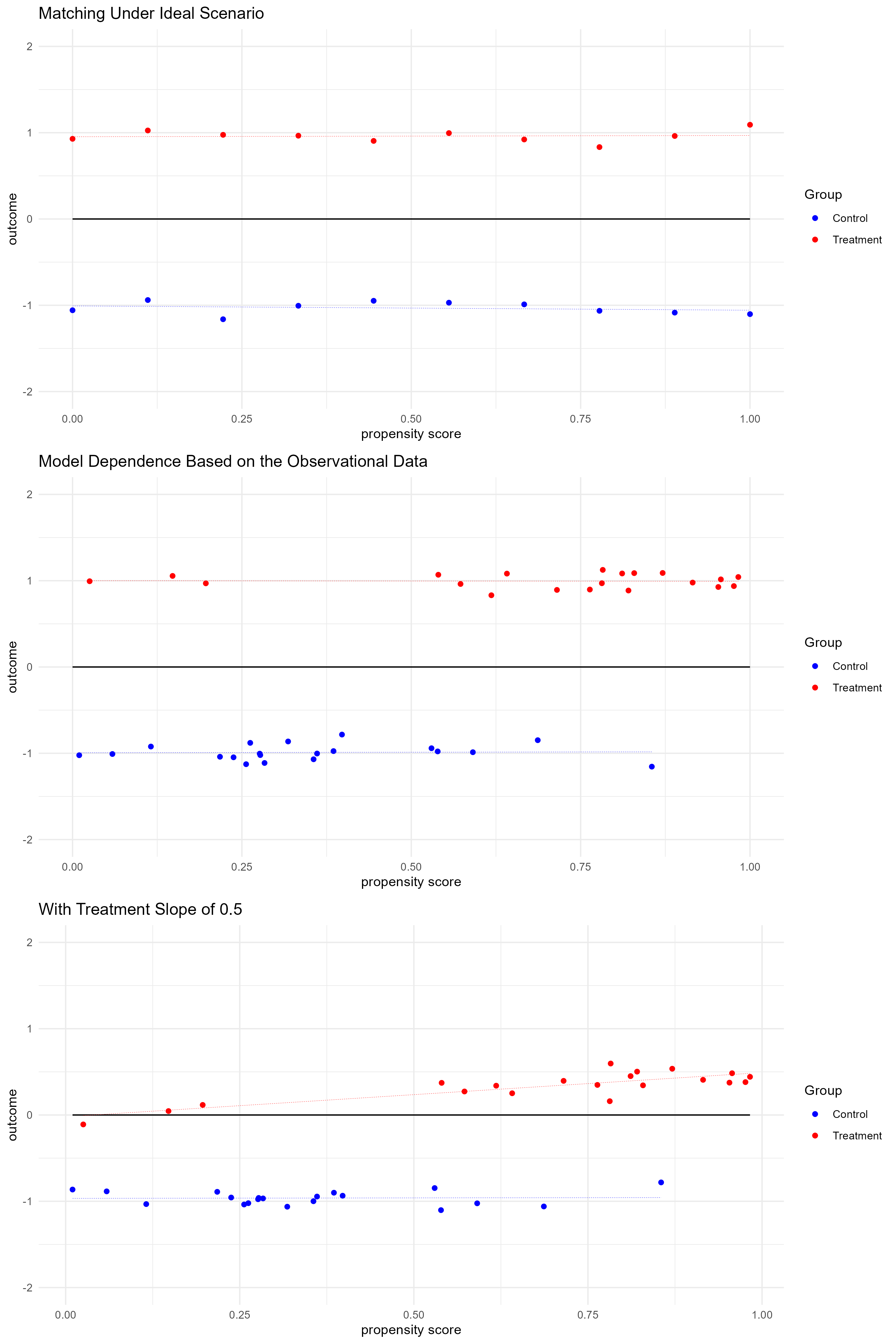}
    \caption{Illustrations on propensity score matching}
    \label{fig:psm}
    
    \raggedright \small \textbf{Note:} The upper panel shows the ideal scenario for propensity score matching, where for each individual in the treatment group, we could find a corresponding individual in the control group with the same propensity value and achieve one-on-one matching. However, as the lower two panels show, with observational data to train and predict the propensity function, the distributions of the treatment and control individuals are unequal along the propensity score, making the one-on-one ideal matching infeasible. But if for the treatment group and for the control group, the outcome is irrelevant to the propensity scores (the middle panel), using any method to get \(\hat{\pi}_a\) yields the same unbiased estimation for \(\pi_a\). Otherwise, as the lowest panel indicates, the choice of \(\hat{\pi}_a\) yields bias. 
\end{figure}

The problem of PSM, as King and Nielsen (\citeyear{king2019}) advocate, is not on the process of causal inference (from \(\pi_a\) to \(\psi^*_a\), as Equation \ref{unconfound} suggests the unbiased process), nor on the process of statistical inference \(\hat{\pi}_a^k\) to \(\pi_a\) (where \(k\) denotes the \(k-\)th method for propensity estimation, and all \(\hat{\pi}_a^k\) can be an unbiased estimator for \(\pi_a\) if causal assumptions Assumption \ref{ciass} hold), it lies in the choice of \(\hat{\pi}_a^k\), or the problem they call "model dependence": we rely on empirical observational data to simulate the DGP for propensity function, and further use the simulated function to predict the propensities for individuals from treatment and control groups (the dots in Figure \ref{fig:psm}), leaving cases unmatched due to uneven densities for the treatment and control groups. The bias, as King and Nielsen (\citeyear{king2019}) suggest, is a \textit{subjective} bias originating from model choices: and the subjective choice of model somehow increases the imbalance\footnote{Imbalance refers to the derivations to the exact match.}, model dependence, and bias for the causal estimation.  \\

Intrinsically, the problem with PSM, if any, still results from the models to specify propensity function cannot satisfy the unconfoundedness assumption. Suppose the unconfoundedness assumption holds, which specifies the independence between the outcome and the treatment under the covariates. In that case, we can imagine that the expected outcomes for the treated and the control should have a constant distance along the propensity score (the slope does not necessarily have to be zero though). Thus, as long as we have a balanced match along the propensity score, the choices of \(\hat{\pi}_a^k\) would be the same and unbiased estimator for \(\pi_a\) (see the illustration of the middle and lower panel of Figure \ref{fig:psm}). However, since the remaining covariates still influence the distribution of outcomes for treatment and control groups based on the propensity scores (for example, in the lower panel of Figure \ref{fig:psm}, we have different slopes for treatment and control groups along the x-axis), and different matching criteria result in bias.\\

In summary, weighing and matching methods achieve causal inference by identifying the propensity function: \(\pi_a(X) = P[A = a |X]\) and expect the estimated \(\hat{\pi}_a(X)\) is an unbiased estimator for \(\pi_a(X)\). Unlike the local average treatment effect models, which restrict the exogeneity for the covariates (discussed below), we do not have any requirements for \(X\) in relation to \(Y\). In other words, the propensity function only matters for \(X\) and \(A\)\footnote{Strictly speaking, the unconfoundedness assumption in the propensity function based method turns to: the outcome is independent to the treatment conditioned on the propensity score, \((Y(1), Y(0)) \indep A \mid \pi_a(X)\). }, and it is almost unavoidable that we have statistical error between the estimated propensity score and the true value (similar to the omitted variable bias in regression analysis). \\

Thus, different models that researchers adopt will unavoidably generate different sample estimators. So, how do researchers claim that the causal effect they captured makes sense by adopting a specific weighting or matching method? We believe two things researchers need to claim before they describe their causal findings: one is the preconditions the models rely on: for instance, what covariates they have included and how they contribute to address or reduce the bias from confounding effects; the other is the theoretical guidance for them to choose the preconditions: the choice of the specific causal identification with variables included would be better theory-driven than pure data-driven. \\

Methods based on propensity functions with weighting and matching have multiple variant forms other than IPW and PSM. For instance, researchers could adopt an evolutionary search algorithm \citep{diamond2013} or a Hungarian algorithm \citep{rosenbaum1989} to optimize the matching process, or stratify the samples and match within different stratification \citep{rosenbaum1984}, or match cases based on Mahalanobis distance (MDM, see \citeauthor{rubin1980} \citeyear{rubin1980}; \citeauthor{king2019} \citeyear{king2019}). 

    \item \textit{Local Average Treatment Effect}\\
    
   The unconfoundedness assumption does not require the full model predicting the propensity for the treatment; indeed, we need the conditions "isolating" the treatment and outcome relationship. We regard the causal relationship between the treatment and the outcome as the \textbf{endogenous effect} (in the isolated system), and the endogenous treatment effect is actually the \textbf{local average treatment effect (LATE)} \citep{imbens1994}. The LATE captures the causal effect by nullifying the exogenous variation. According to Equation \ref{expectation}, with no exogenous variability, the outcome conditioned on specific treatment status equals the potential outcome (i.e., \(E[Y \mid A = a] = Y(a)\)). Classical \textbf{instrumental variables (IV)} method uses the exogenous instrumental variables, and \textbf{fixed effect (FE)} method focuses on the endogeneity effects within entities (groups, individuals, etc.). We discuss them briefly in this subsection. Meanwhile, as we suggest in the weighting and matching method, to specify the causal effect with the LATE method, we suggest researchers point out what the precondition, and in this case, the exogeneity condition, is for the inference.
    \begin{itemize}
        \item \textit{Instrumental Variables}\\
        We start with the assumption of the IV method. A comprehensive review from the biostatistical and clinical perspective of the IV can be found in \citeauthor{baker2016} \citeyear{baker2016}. Specifically, we illustrate a binary IV scenario for simplifcation. In an empirical study, suppose we have assigned individuals to the treatment and control groups to measure the post-treatment divergence as the treatment effect. We have discussed that the main concern in this setting is the action of "assignment"-- whether it is "random." With random assignment, as we have discussed, 
        \begin{align*}
            &\hat{\psi} = E[Y_i | A_i =1 ] - E[Y_i | A_i = 0] \\= &\psi = E[Y | A =1 ] - E[Y | A = 0] \\= &\psi^* = E[Y(1)]-E[Y(0)],
        \end{align*}
     as the fundamental of causal inference with statistical estimands. \\
     
     Now that we consider some "naughty" individuals do not want to follow the group assignment in the process described above, violating the randomization in the grouping. Let \(X\) denote the binary status of whether they are assigned to the treatment group, and \(A\) denote whether they receive treatment. We then have four observational groups: assigned treatment, received treatment: (\(X_i = 1, A_i = 1\)); assigned treatment, received control: (\(X_i = 1, A_i = 0\)); assigned control, received treatment: (\(X_i = 0, A_i = 1\)); and assigned control, received treatment: (\(X_i = 0, A_i = 0\)). The naughty individuals who choose whether to comply with the treatment assignment are either random or affect the outcome only through treatment \footnote{This is not to say that the "naughty decisions" are not affected by other covariates. For instance, some individuals care about their past medical history and decide not to take the treatment. Researchers do not care about the covariates: as long as there's no other reasonable correlation with the outcome except via the treatment.}. \(X\) are the instruments, with the (instrumental) exogeneity assumption\footnote{The general exogeneity assumption suggests the explanatory variables should be uncorrelated with any error term: \(cov(\epsilon,X) = 0\), as \(\epsilon = Y - f(X)\) if \(f(X)\) is the predicted value for the outcome given covariates are \(X\).}:\\
     \begin{ass}[Instrumental Exogeneity Assumption]\label{exogeneity}
         The instrumental variable selection needs to satisfy the following two conditions:\\
         a) Instrument relevance: he instrumental variable should be correlated with the treatment: 
         \[
         cov(X, A) \ne 0
         \]
         b) Instrument Exogeneity: the instrumental variable should be uncorrelated with the outcome fitted with the treatment:
         \[
         cov(X, Y|A) = 0
         \]
     \end{ass}
     
    For each of the four observational groups, we may guess their motivations for complying with or violating the groups they are assigned to. For instance, individuals complying with the treatment group may be \textbf{always-takers}, as they intend to take the treatment no matter which group they are assigned, or \textbf{compliers}, who just comply with their assignment, while individuals violating the treatment group assignment may be \textbf{defiers}, who simply go to the opposite group they are assigned, or \textbf{never-takers}, who intend to go to the treatment group no matter which group they are assigned. Similarly, those who comply with the control group are either compliers or never-takers, and those who are assigned control but go to the treatment group are either defiers or always-takers. In other words, always-takers have individuals with (\(X_i = 1, A_i = 1\)) and (\(X_i = 0, A_i = 1\)), compliers are individuals with (\(X_i = 0, A_i = 0\)) and  (\(X_i = 1, A_i = 1\)), never-takers are those (\(X_i = 0, A_i = 0\)) and  (\(X_i = 1, A_i = 0\)), and defiers are those  (\(X_i = 0, A_i = 1\)) and  (\(X_i = 1, A_i = 0\)). \\

    It is easy to notice that we can not identify the causal effects for always-takers and never-takers with the instrument, as the treatment effect is not affected by the instrument. Further, we need to assume that there are no defiers in the experiment, as we need the relationship between the treatment and the instrument monotonic: 
    \begin{ass}[Monotonic Assumption for Instrumental Variables]\label{monotonic}
         There are no defiers in the instrumental settings. In other words, the instruments affect the treatment in one direction (either always increases or always decreases the likelihood of receiving the treatment).
    \end{ass}

Thus, the treatment effect on the outcome for the compliers can be reliably estimated using the instrument, provided the exogeneity and monotonicity assumptions hold. Since adding the instrument meets all the assumptions necessary for causal inference, particularly unconfoundedness, the outcome is now independent of the treatment given the instrument. The Local Average Treatment Effect (LATE) for the compliers is determined by dividing the covariance between the outcome and the instrument by the covariance between the treatment and the instrument.
    \begin{equation}\label{iv}
        LATE = \frac{cov(X, Y)}{cov(X, A)} = \frac{E[Y| X = 1]- E[X = 0]}{E[A | X = 1]- E [A |X=0]}
    \end{equation}
    for the binary instrument. We call the generalized form (the division for the two covariances) as the IV estimand \footnote{IV method first appears in econometric literature not for identifying the causal effects or the endogeneity. It aimed to address the parameter identification in \textbf{simultaneous equation models}, in which the outcome is the equilibria of some structural relationships (we call these equilibria the \textbf{structural form}). For instance, consider the outcome as an equilibrium of the price and the quality, say \(Y = f(P, Q)\). We have the equilibrium: \(Q_s = Q_d\), where \(Q_s\) and \(Q_d\) represent the quality from the supply and the demand. Suppose we have the supply function (in linear form, with intercept excluded), and the demand function:
    \[
    \begin{cases}
        P_s = f(Q_s, X_s) +\epsilon_s = \alpha_s Q_s + \beta_s X_s + \mu_s \\
        P_d = f(Q_d, X_d) +\epsilon_d = \alpha_d Q_d + \beta_d X_d + \mu_d
    \end{cases}
    \] where \(X_s, X_d\) are separately the covariates affecting the relationship between price and quality for supply and demand (they are more likely different sets). With equilibrium conditions, \(Q\) and \(P\) are endogenous variables (since they appear in both structural forms), and we put the endogenous variables on the left side, while the exogenous variables on the right, so we have:
    \[
    \begin{cases}
        P =  \pi_1 X_s + \pi_2 X_d + \nu_p \\
        Q =  \theta_1 X_s + \theta_2 X_d + \nu_q
    \end{cases}
    \]
    which are the \textbf{reduced forms}. With some algebra transformation, we could easily get: 
    \[
\pi_1 = \frac{-\beta_d}{\alpha_d - \alpha_s}; \pi_2 = \frac{\beta_s}{\alpha_d - \alpha_s};
\theta_1 = \frac{-\alpha_s \beta_d}{\alpha_d - \alpha_s};
\theta_2 = \frac{\alpha_d \beta_s}{\alpha_d - \alpha_s}\]\[
\alpha_s = \frac{\theta_1}{\pi_1};
\alpha_d = \frac{\theta_2}{\pi_2};\beta_s = \pi_2(\frac{\theta_1}{\pi_1}-\frac{\theta_2}{\pi_2});
\beta_d = -\pi_1(\frac{\theta_1}{\pi_1}-\frac{\theta_2}{\pi_2})
\]
so, for example, if our target is the slope in the demand curve between \(Q\) and \(P\): \(\alpha_d = \frac{\theta_2}{\pi_2}\), we need the auxiliary variable \(X_d\) and \(X_s\) satisfying \(X_d \indep X_s\)(otherwise, the estimation on \(\theta_2\) and \(\pi_2\) are biased as we omitted the variables in the reduced form) a to establish models separately for \(Q\) and \(P\), so that we get the unbiased estimator on the slope. The tricky thing is, in the simultaneous structural model for observational data, \(X_d\) and \(X_s\) will overlap; thus, in reality, we cannot differentiate \(\theta_1\) with \(\theta_2\), and \(\pi_1\) with \(\pi_2\), so we will doubtlessly yield a biased estimator. \\
So the solution is to deploy an instrument and to circumvent direct estimation on the slope for the demand function; rather, we find the exogenous function which identifies \(Q_s\) using the variable has nothing to do on the demand side, then rely on \(Q_d = Q_s\) to replace the quantity in demand with the quantity identified on the supply side, and finally calculate the slope between \(Q_s\) and \(P_s\). For example, consider we are describing the demand curve between the price of grains and the quantity. We introduce the instrument, the rainfall, which affects only the quantity in supply instead of demand, and with the steps described, we can identify the demand curve slope \citep{pearl2015, haavelmo1944, stock2003}. With the exogenous variable uncorrelated with the error term in the output model, we address the consistent and unbiased correlation between the endogenous variables. \\
The process discussed above reveals the essence of the instrument variable lies in the unidentifiable problem in matrix operation. Let's further simplify the simultaneous structural model as: \[\begin{cases}
    Y_{n \times 1} = X_{n\times m}A1_{m \times 1} + Z_{n \times t}B1_{t \times 1} + e1_{n \times 1} \\
    X_{n \times m} = Y_{n \times 1}A2_{1 \times m} + Z_{n \times t}B2_{t \times m} + e2_{n \times m}.
\end{cases}\] In the simultaneous structural model, \(X\) and \(Y\) are endogenous variables, while \(Z\) is exogenous. The suffix denotes the dimension of the matrix (we suppose there are \(n\) cases, \(m\) endogenous covariates, and \(t\) exogenous covariates (assume \(t \le m\)). To estimate the covariates, our first step is to bring the second equation directly to the right of the first equation, simplifying as \( Y_{n \times 1} = Y_{n \times 1}A2_{1 \times m}A1_{m \times 1} + Z_{n \times t}(B2_{t \times m}A1_{m \times 1}+B1_{t \times 1}) + (e2_{n \times m}A1_{m\times 1}+ e1_{n \times 1})\). Let \(A = A2A1, C = (B2A1+B1)(I - A)^{-1}\), We thus capture the reduced form: \(Y_{n \times 1} = Z_{n \times t}C_{t \times 1} + e\). Since \(C\) is a \(t \times 1\) vector, we could not calculate specifically \(A1, A2, B1, B2\), unless we make some assumptions that some vectors in the matrices are 0. This is why we need the exogeneity assumption to presume further some uncorrelated relationship between the instrument and the error term to get the estimation in the endogenous relationship.}. The pseudo-randomization is achieved with the assistance of the instrument for the compliers.\\

A classic example in social science is to measure the causal effect of military service on future earnings \citep{angrist1990}. Since the endogeneity of voluntary enlistments, individuals who choose to enlist may differ from those who do not, the causal relationship estimated directly from the model between military services and earnings yields bias. To address the endogeneity issue, Angist uses the draft lottery during the Vietnam War as the instrument: the draft lottery numbers are randomly assigned, while the eligibility to be drafted influences the likelihood of military service. Thus, those who are and are not induced to serve due to the draft lottery are the compliers. For the compliers, the study showed that those with military service experiences have lower future earnings than those without, so it verifies the negative causal effect of military service on future earnings. \\

Does this study have a good research design? First, we need to point out the external validity of the causality: in this study, Agrist focuses on the draft between 1970 and 1972 (so there are only three years for the draft) and men born between 1950 and 1953. So, rigorously speaking, the result can only be applied to the very specific men group in the 1950-53 cohort. Second, since many people resist being drafted, does draft evasion invalidate the results? Not at all! This is because the draft lottery is an exogenous variable, so if the results are restricted to the compliers. But, this relies on the monotonic assumption, which is to say that those drafted but did not serve in the military are never-takers: they determined that whatever the results are, they will not serve in the military-- in other words, they are not defiers. Under this circumstance, the results are valid for the compliers-- the average treatment effect yields between those drafted and did military service, and those not drafted and didn't go to the military. \\

In today's social science, "finding an appropriate instrumental variable is rather an art than a science." This is because having a strictly exogenous variable in observational survey data is hard and sometimes even requires imagination. Traditionally, social scientists rely on some naturally random assignments \citep{angrist2001}, for instance, geographic or spatial variables \citep{card1999} and weather variables \citep{dell2009}, or deliberately engineered changes, for instance, policy changes \citep{angrist1999} and economic shocks \citep{autor2013}, and even demographic \citep{oreopoulos2006}, biological and health \citep{fletcher2009}\footnote{When using demographic, biological, and health variables as the IV, the core research questions usually set on treatment effect of a policy on the socioeconomic outcome, while when using policy and economic shocks, the research questions are usually the relationship between socioeconomic variables.} as the instrumental variables in their studies. \\

Finally, an unavoidable situation for social scientists is that the variables they have to rely on are \textbf{weak instrumental variables} \citep{stock2002}\footnote{The weak instrumental variable issue discusses more about the statistical error between the estimator from the observational data and the statistical estimand, which we will mainly focus on Section \ref{sec3}. We decide to put it here to ensure coherence in the discussion of the IV method.}. According to Assumption \ref{exogeneity}, a valid instrument should satisfy the conditions of both relevance and exogeneity, a weak instrument is defined as its relevance with the treatment is weak. Therefore, the instrument does not explain much of the variation in the treatment, which is exogenous, leading to two consequences: first, we may not reliably estimate the LATE due to endogeneity, and our estimation will be biased towards the model which we didn't use any instrument (compared to a strong IV); second, with less precise estimates on the LATE, the confidence interval of our estimation will be wider. To illustrate this, consider the identification models between the outcome and the treatment, and between the treatment and the instrument:
\[
\begin{cases}
    Y = g(A) +\epsilon \\
    A = h(X) + \eta 
\end{cases}
\]
In IV settings, the operation we estimate the LATE for the compliers is a two-step model (if we set the model in a parametric way, we use two ordinary least square models, and this is called the \textbf{two-stage-least-square (2SLS)}). We first estimate \(\hat{A} = \hat{h}(X)\), and put the predicted \(\hat{A}\) in the first equation to yield the IV estimator \(\hat{g}_{IV}(A) = E[Y \mid \hat{A}]\). If the IV is correctly identified, we have \(\hat{A} \to h(X)\), meanwhile \(E[\epsilon \mid X = 0]\). Therefore, 
\begin{equation} \label{iveq1}
\hat{g}_{IV}(A) =  E[Y \mid \hat{A}] \to E[Y \mid h(X)] = E[g(A) + \epsilon \mid h(X)] = g(A)    
\end{equation}

According to the definition of weak IV, which has a weak relevance with the treatment, we may assume that \(\hat{A} = \hat{h}(X) \to 0\). Still, \(E[\epsilon |X = 0]\). Therefore, 

\begin{equation} \label{iveq2}
\begin{aligned}
\hat{g}_{IV}(A) &= E[Y \mid \hat{A} \to 0] = E[Y \mid \hat{A} \to 0] = E[g(A) + \epsilon \mid \hat{A} \to 0]   \\& =E[g(A) \mid \hat{A} \to 0] + \underbrace{E[\epsilon \mid \hat{A} \to 0]}_{:=0} = E[g(E[A]) \mid \hat{A} \to 0] = g(E[A])     
\end{aligned}
\end{equation} 
Therefore, the estimator from the weak IV approaches to the result of purely the second-stage model in the two-stage estimation if the relevance between the instrument and the treatment approaches zero. Combine Equations \ref{iveq1} and \ref{iveq2}, we have the bias from weak IV, which is:
\[
\text{Bias} = g(E[A]) - g(A). 
\]
    \item \textit{Fixed Effects}\\
    
    As discussed above, the essence of the LATE method is to isolate the exogenous variation in the endogenous variables (the treatment) and obtain an unbiased and consistent estimator for the treatment effect. The fixed effect method could also be used to make the unconfoundedness assumption hold by isolating the exogenous variation, as it focuses on the variations within the entity. We assume the values of the outcome variable follow a normal distribution: 
    \[
    y_{ij} \sim N(\alpha_j + f(a_{ij}), \sigma^2_y)
    \]
    Suffix \(i\) refers to the individuals, or the granular level of data, while suffix \(j\) refers to the group unit, which aggregates the individual-level data. \(alpha_j\) denotes the group-level specific term, which accounts for the variability in the mean response \(y_{ij}\) between different groups \(j\).\( f(a_{ij}) \) represents the effect of the treatment \( a_{ij} \) on the response variable \( y_{ij} \), and \(\sigma^2_y\) denotes the variance for the outcome. We could further rewrite the group-level term as\footnote{The assigned probability distribution for \(\alpha_j\) in multilevel models is another "soft constraint" in this type of models (see \citealt[ch. 12, p. 257]{gelman2006}).}:
    \[
    \alpha_j \sim N(\mu, \sigma^2_\alpha)
    \]
    In which \(\mu\) denotes the overall mean effect across all groups, and \(\sigma^2_\alpha\) represents the variance of the group-specific effects\footnote{the representation is equivalent to the linear form expression: \(y_ij = \mu + f(a_ij) + \epsilon_{\alpha_j} + \epsilon_{y_{ij}}\), in which we assume the error term\(\epsilon_{y_{ij}}\) is uncorrelated with the treatment \(a_ij\), group-level fixed mean \(\mu\), and the group-level error term \(\epsilon_{\alpha_j}\).}. Since we set up the distribution for a normal, we have specified the unrelated relationship between the treatment effect \(f(a_{ij})\), the group-level effect \(\alpha_j\), and the statistical error (variance) of the outcome \(\epsilon_{ij}\) for any given group \(j\) ( i.e., \(E[\sigma_{y} \mid f(a_{i1}),f(a_{i2}),..., f(a_{iJ}),\alpha_j] = 0\)). Also, we assume that the error terms between any two groups are uncorrelated: \(E[\sigma_{y_{it}} \mid \sigma_{y_{is}}] = 0 \text{ if } t \ne s\). \\

    With the assumptions hold, \(f(a_{ij})\) in the model denotes the LATE for individuals. For instance, consider our aim is to estimate an individual's education \(a_{ij}\) on her earnings \(y_{ij}\) from the panel data. Therefore, the group \(j = 1, 2, ..., J\) represents the specific waves, and we could simply calculate the ATE of education on earnings simply by demeaning the outcome:
    \[
    (y_{ij} - \bar{y}_i) = (f(a_{ij}) - f(\bar{a}_i)) + (\sigma_y - \sigma_\alpha)
    \]

    In the expression above, \(\bar{y}_i\) and \(\bar{a}_i\) separately represent the mean outcome and treatment for the specific individual. Since the exogenous variables (i.e., time-invariant individual characteristics, ability and intelligence, and initial conditions, like early childhood health) are controlled by the fixed term, we could make inferences on our treatment and the outcome, supposing there are no time-varying variables affecting the results (which is restricted by \(E[\sigma_{y_{it}} \mid \sigma_{y_{is}}] = 0\) in our model setting). \\

    As can be seen from the above example, if we set the group level as time, the fixed effect is indeed the same as the difference-in-difference model that we discussed before. In other words, DID is a specific application of the FE under the panel data settings. We discussed the advantage of DID in addressing the violation of the positiveness assumption, and indeed, it addresses the violation of the unconfoundedness assumption, simply by ignoring the exogeneity fixing the time-invariant effects. 
    \end{itemize}    
\end{itemize}
\section{Analyzing the Statistical Error}\label{sec3}
In the section above, we mainly discussed the methods addressing the causal error, in other words, how we use the statistical estimand to approximate the causal estimand \(\psi- \psi^*\). We mentioned in passing some content of this section, which is how to estimate from the statistical estimator from the observational data to approximate the statistical estimand: the statistical error \(\hat{\psi} - \psi \). According to Equation \ref{staterror}, the statistical error can be decomposed in the statistical variance \(\hat{\psi} - E[\hat{\psi}]\) and the statistical bias \(E[\hat{\psi} - \psi]\). In (especially) machine learning literature, we often discuss a balance to be struck between minimizing the bias or minimizing the variance to achieve optimal model performance, and the balance is called \textbf{variance-bias tradeoff} \footnote{We will simply find we could express the mean square error (MSE) of the statistical error by the statistical variance and the statistical bias: the Mean Squared Error (MSE) of an estimator \(\hat{\psi}\) is defined as:
\(\text{MSE}(\hat{\psi}) = E[(\hat{\psi} - \psi)^2]\). We add and subtract the expected value of the estimator \(E[\hat{\psi}]\) inside the squared term:
\[
\text{MSE}(\hat{\psi}) = E[(\hat{\psi} - E[\hat{\psi}] + E[\hat{\psi}] - \psi)^2]
\]

Then expand the squared term:
\[
\text{MSE}(\hat{\psi}) = E[(\hat{\psi} - E[\hat{\psi}])^2 + 2(\hat{\psi} - E[\hat{\psi}])(E[\hat{\psi}] - \psi) + (E[\hat{\psi}] - \psi)^2]
\]

Simplify by recognizing that \(E[\hat{\psi} - E[\hat{\psi}]] = 0\) (since the expectation of a deviation from the mean is zero):
\[
\text{MSE}(\hat{\psi}) = E[(\hat{\psi} - E[\hat{\psi}])^2] + 2E[\hat{\psi} - E[\hat{\psi}]](E[\hat{\psi}] - \psi) + (E[\hat{\psi}] - \psi)^2
\]
\[
\text{MSE}(\hat{\psi}) = E[(\hat{\psi} - E[\hat{\psi}])^2] + (E[\hat{\psi}] - \psi)^2
\]

Recognize that \(E[\hat{\psi} - E[\hat{\psi}]] = 0\) means the middle term drops out:
\[
\text{MSE}(\hat{\psi}) = \text{Var}(\hat{\psi}) + \text{Bias}(\hat{\psi})^2
\]

Thus, the decomposition of the Mean Squared Error (MSE) is:
\[
\text{MSE}(\hat{\psi}) = \text{Var}(\hat{\psi}) + \text{Bias}(\hat{\psi})^2
\].}. Although some biased estimators might be useful as they could optimize the model with a minimal \textbf{mean square error (MSE)}\footnote{A common case for a biased estimator is to estimate the variance of a normal distribution, which yields smaller MSE than use the unbiased estimator. Given a set of independent and identically distributed (i.i.d.) samples \(X_1, X_2, \ldots, X_n\) from a normal distribution \(N(\mu, \sigma^2)\), we can estimate the population variance \(\sigma^2\). The sample variance \(S^2\) is an unbiased estimator of the population variance \(\sigma^2\). It is defined as:
\[
S^2 = \frac{1}{n-1} \sum_{i=1}^n (X_i - \bar{X})^2,
\]
where \(\bar{X}\) is the sample mean:
\(
\bar{X} = \frac{1}{n} \sum_{i=1}^n X_i.
\).
The sample variance \(S^2\) satisfies:
\(
E[S^2] = \sigma^2.
\).

A biased estimator of the population variance is the biased sample variance, defined as:
\[
\sigma^2_{\text{biased}} = \frac{1}{n} \sum_{i=1}^n (X_i - \bar{X})^2.
\]

This estimator is biased because:
\(
E[\sigma^2_{\text{biased}}] = \frac{n-1}{n} \sigma^2.
\)
Despite its bias, \(\sigma^2_{\text{biased}}\) can be preferred (for instance, in industrial manufacturing). This is because, in some cases, the latter is easy to compute, while we do not need a precise estimation of the overall variance. Plus, the reduction in variance provided by the biased estimator can outweigh the increase in bias (especially in smaller sample cases), leading to a lower overall MSE.}, in causal inference methods discussed in the whole thesis, when we calculate the statistical estimator, we require it to be an unbiased estimator on the statistical estimand, so we have \(E[\hat{\psi}] - \psi \equiv 0\).\\

The analytical tool on how our estimator can infer the estimand unbiased is the \textbf{asymptotical analysis}. We imagine our estimator \(\hat{\psi} = \psi(\mathbb{P}_n)\) is a nuisance perturbed estimation on the true estimand \(\psi(\mathcal{P}\). Thus, we denote the perturbed distribution as \(\psi(\tilde{\mathcal{P}}_\epsilon)\big|_{\epsilon = 0}\), where \(\epsilon\) denotes the infinitesimal error term which approaches zero. Our core questions are two folds: 1) to which degree \(\tilde{\mathcal{P}}_\epsilon\) drifted from the true measure \(\mathcal{P}\); and 2) how the perturbation in the distribution of \(\tilde{\mathcal{P}}_\epsilon\) affects the estimation of \(\hat{\psi}\) on \(\psi\). We use two functional forms to identify the two questions: we use the score function to determine the direction of the drift in measure, and we use the influence function to identify the effect of the perturbation. As we will elaborate on this section, from the perspective of asymptotical analysis, the best causal estimator we will derive is the unbiased efficient estimator under the condition of regularity and asymptotical linearity: this estimator is not only unbiased to the statistical estimand, but also yields the smallest variance. We will show below that the efficient estimator is also robust (Neyman orthogonal) from the econometric perspective, and the influence function is also efficient. \\

Due to the unfamiliarity of the materials discussed in this section for readers, we arrange it into the following parts: we will first introduce the characteristics of the regular and asymptotically linear estimators, and we further discuss the detailed characteristics of the score function and the influence function, revealing their connections and how they assist us in finding the efficient estimator. Finally, we derive the efficient influence function and the efficient estimator for the average treatment effect. In the following chapters of this thesis, we will still use the mathematical equations in this section to derive different efficient estimators adapting to various social science and demographic research scenarios. 

\subsection{Introduction to RAL Estimators}
For all the RAL estimators, it would be best to find the one with the lowest variance so that the MSE for the estimator towards the statistical estimand would be lowest. Fortunately, for the unbiased estimator, we know the lowest bound for the variance, given by the \textbf{Cramer-Rao bound}:
\begin{lem}[Cramer-Rao Bound]\label{crb}
     The Cramer-Rao Bound (CRB) states that for an unbiased estimator \(\hat{\psi} = \psi(\mathbb{P}_n)\) of \(\psi = \psi(\mathcal{P})\), the variance of \(\hat{\psi}\) is at least as large as the inverse of the Fisher information:

\[
\text{Var}(\hat{\psi}) \geq \frac{1}{I(\psi)}
\]

where \(I(\psi)\) is the Fisher information given by:

\[
I(\psi) = \mathbb{E} \left[ \left( \frac{\partial}{\partial \psi} \log f(Z; \psi) \right)^2 \right]
\]

\end{lem}

The proof will be given later in this section. For all the unbiased estimators, if the variance attains the CRB, we call the estimators the \textbf{efficient estimators}. However, as we will elaborate, efficient estimators are sometimes hard to derive due to the complexity of the underlying functional form (mostly the influence function; see below), and we need deterministic or heuristic approaches that will make our estimator more efficient than any other peers we may find. \\

Besides unbiased, In our study, due to the central limit theorem (Theorem \ref{clt}), we further require the estimator to be \textbf{regular and asymptotically linear (RAL)}. \textbf{Regularity} in general means that our estimator processes towards some desirable characteristics as the sample size increases: 
\begin{equation}\label{regu}
    \sqrt{n}\big(\psi(\mathbb{P}_n) - \psi(\mathcal{P})\big)\stackrel{\mathcal{P}}{\rightsquigarrow}\mathcal{D},
\end{equation}
Where \(\mathcal{P}\) denotes an empirical measure which we will discuss shortly, and \(\mathcal{D}\) denotes a fixed distribution\footnote{In this thesis, the denotation on the convergence uses the expressions in \citeauthor{vandervaart1998} (\citeyear{vandervaart1998}). In short, convergence in distribution (weak convergence) \(X_n \xrightarrow{\mathcal{D}} X \iff X_n \rightsquigarrow X \) suggests that a sequence of random variables \( \{X_n\} \) converges in distribution to a random variable \( X \) if for all points \( t \) at which \( F_X(t) \) is continuous:\( \lim_{n \to \infty} F_{X_n}(t) = F_X(t), \)
where \( F_{X_n}(t) \) and \( F_X(t) \) are the cumulative distribution functions of \( X_n \) and \( X \); convergence in probability \( X_n \xrightarrow{\mathcal{P}} X \) (or \( X_n \xrightarrow{prob.} X \), to differentiate with the measure \(\mathcal{P}\))  suggests the relationship between \(X_n\) and \(X\), for every \(\epsilon > 0\), is: \(\lim_{n \to \infty} P(|X_n - X| \geq \epsilon) = 0\). Almost sure convergence \( X_n \xrightarrow{a.s.} X \) suggests \(X_n\) almost surely converges to \(X\)\( P\left( \lim_{n \to \infty} X_n = X \right) = 1\). Meanwhile, we also use big-O probability and small-o probability to denote convergence: big \(O_p: X_n = O_p(a_n)\) indicates that the sequence of random variable \(X_n\) is bounded by \(a_n\) in probability: for every \(\epsilon > 0\), there exist constants \(M > 0\) and \(N > 0\) such that\(
P(|X_n| \leq M a_n) \geq 1 - \epsilon \quad \text{for all } n \geq N.
\) Small \(o_p\) suggests that \(X_n\) is asymptotically smaller than \(a_n\) as \(n\) increases, or the difference between \(X_n\) and \(a_n\) is negligible: for every \(\epsilon > 0\) and \(\delta > 0\),\(
P(|X_n| \geq \delta a_n) \to 0 \quad \text{as } n \to \infty.\)}. \textbf{Asympotitical linearity} suggests the difference between the estimator and the statistical estimand can be approximated by the linear combination of the appropriate normalized sum of the IID random variables. Suppose the function is \(\phi(\psi; \mathcal{P}; Z_i)\), for the estimator \(\psi\), measure \(\mathcal{P}\) and dataset \(Z_i = (X_i, Y_i)\)\footnote{The influence function has three entries, \(\phi\) refers to the estimation functional form, \(\mathcal{P}\) refers to the measure. \(Z_i\) refers to the measurable set (dataset). Due to our assumptions in Section \ref{sec1} that the estimation function and the measure (DGP) will not change simultaneously, thus, if we specify the influence function for an estimator, we omit the measure and the dataset. We use \(Z_i\) for discrete elements in the measurable set and \(z\) if the elements are continuous.},therefore, 
\begin{equation}\label{AL}
\sqrt{n}\big((\psi(\mathbb{P}_n) - \psi(\mathcal{P}))- \frac{1}{n}\sum_{i=1}^n\phi(\psi; \mathbb{P}_n;Z_i)\big) \xrightarrow{prob.} 0,
\end{equation}
We call \(\phi(\psi;  \mathbb{P}_n; Z_i)\) as the \textbf{influence function} for the empirical data \(Z_i\) with respect to the empirical measure \(\mathbb{P}_n\). With Theorem \ref{clt}, asymptotical linearity could also be expressed as:
\begin{equation} \label{AL2}
\sqrt{n}(\psi(\mathbb{P}_n) - \psi(\mathcal{P})) \xrightarrow{d} \mathcal{N}(0, \sigma^2).    
\end{equation}

where \(\sigma^2 = \mathrm{Var}(\phi(\psi; \mathbb{P}_n; Z_i ))\) and we call it the \textbf{asymptotic variance}. Equation \ref{AL} can also be expressed as:
\begin{equation}\label{AL3}
    \big(\psi(\mathbb{P}_n) - \psi(\mathcal{P})\big)- \frac{1}{n}\sum_{i=1}^n\phi(\psi; \mathbb{P}_n; Z_i) = o_p(n^{-1/2}) 
\end{equation}
which stresses the approximation error (remainder term) becomes negligible faster than \(n^{-1/2}\)in probability. The property revealed in Equation \ref{AL3} is very useful when we try to yield the \textbf{doubly robust/debiased machine learning} estimator (in some literature, we also denote \(R_2(\mathbb{P}_n; \mathcal{P}) = \big((\psi(\mathbb{P}_n) - \psi(\mathcal{P}))- \frac{1}{n}\sum_{i=1}^n\phi(\psi;  \mathbb{P}_n; Z_i)\big)\) and call it as the \textbf{second-order remainder})\footnote{Not all unbiased estimators are RAL. For instance, consider the median estimator for all symmetric distributions. It is an unbiased estimator for the population median, but it is not asymptotically linear as \(\phi(\psi; Z)\) for the median is not a smooth function. Another classic example here is the Hodges' estimator.}. We will discuss regularity and asymptotically linearity in detail in the following subsections. 

\subsection{Regularity and Score Function}
We start with the definition of \textbf{the score function}. Suppose we have two (\(\sigma-\)finite) measures \(\mathcal{P}(Z)\) and \(\tilde{\mathcal{P}}(Z)\) where \(\mathcal{P}(Z)\) dominates \(\tilde{\mathcal{P}}(Z)\) (or \(\tilde{\mathcal{P}}(Z)\) is absolutely continuous with respect to \(\mathcal{P}(Z)\)). Now imagine we have a \textbf{differentiable path} starting from \(\mathcal{P}(Z)\) and ending at \(\tilde{\mathcal{P}}(Z)\). The most convenient method to define it is linear interpolation: let \(\tilde{\mathcal{P}}_\epsilon(Z) = \epsilon \tilde{\mathcal{P}}(Z) + (1-\epsilon)\mathcal{P}(Z) (\epsilon \in [0,1]).\) Therefore, suppose the probability density function for \(\mathcal{P}(Z)\) and \(\tilde{\mathcal{P}}(Z)\) are respectively \(p(z)\) and \(\tilde{p}(z)\), then the density function for \(\tilde{p}_\epsilon(z) = \epsilon \tilde{p}(z) +(1-\epsilon)p(z). \) \\
\begin{proof}
    We have the definition of a probability density function \(p(z)\) for a measure \(\mathcal{P}\) such that for any measurable set \(Z\): \[
    \mathcal{P}(Z) = \int_Z p(z)dz
    \]
Thus, 
\[
   \tilde{\mathcal{P}}_\epsilon(Z) = \epsilon \int_Z \tilde{p}(z) \, dz + (1-\epsilon) \int_Z p(z) \, dz =\int_Z \left( \epsilon \tilde{p}(z) + (1-\epsilon) p(z) \right) dz
   \]
And therefore, 
   \[
   \tilde{p}_\epsilon(z) = \epsilon \tilde{p}(z) + (1-\epsilon) p(z).
   \]
\end{proof}

With the differentiable path, we define the score function corresponding to the path from \(\mathcal{P}(Z)\) towards \(\tilde{\mathcal{P}}(Z)\) as the rate \(\epsilon_0\) change of the log-likelihood at the starting point \(\mathcal{P}(Z)\) (the gradient of the log-likelihood function):
\begin{equation}\label{sf}
    s_{\epsilon_0}(z) = \frac{\partial log \tilde{p}_\epsilon(z)}{\partial \epsilon} \bigg|_{\epsilon = \epsilon_0}
\end{equation}
Usually, in our discussion, we set \(\epsilon_0 = 0\). According to Equation \ref{sf}\footnote{Generally, if \(\epsilon_0 = 0\), we simplify the score function\(s_\epsilon(z)\) as \(s(z)\), or further, \(s\).}, we may rewrite \(\tilde{\mathcal{P}}_\epsilon(Z)\) and \(\tilde{p}_\epsilon(z)\) as: \(\tilde{\mathcal{P}}_\epsilon(Z) = \int_Z (1+\epsilon s(z))d\mathcal{P}(Z)\) and \(\tilde{p}_\epsilon(z) = (1+ \epsilon s(z))p\). \\
\begin{proof}
    We first prove that the definition of \(s(z) = \frac{\partial log \tilde{p}_\epsilon(z)}{\partial \epsilon} \bigg|_{\epsilon = 0}\) can be rewrite as: 
\begin{equation}\label{score2}
    s(z) = \frac{\tilde{p}(z)}{p(z)} - 1
\end{equation}
Using the chain rule to differentiate: 
\begin{align*}
\frac{\partial log\tilde{p}_\epsilon(z)}{\partial \epsilon}\bigg|_{\epsilon = 0}  & = \frac{\partial log \tilde{p}_\epsilon(z)}{\partial\tilde{p}_\epsilon(z)}\cdot \frac{\partial\tilde{p}_\epsilon(z)}{\partial \epsilon} \bigg|_{\epsilon = 0}= \frac{1}{\tilde{p}_\epsilon(z)}\cdot \frac{\partial}{\partial \epsilon}(\epsilon \tilde{p}(z) +(1-\epsilon)p(z))\bigg|_{\epsilon = 0} \\& =\frac{1}{p(z)}(\tilde{p}(z) -p(z)) = \frac{\tilde{p}(z)}{p(z)} - 1
\end{align*}
Therefore, \(\tilde{p}(z) = (s(z) + 1)p(z)\). We could rewrite \(\tilde{p}_\epsilon(z)\) as \(\tilde{p}_\epsilon(z) = (1 - \epsilon)p(z) + \epsilon (s(z) + 1)p(z)\). Simplifying the expression:
\begin{equation}\label{linearsf}
    \tilde{p}_\epsilon(z) = \left( 1 - \epsilon + \epsilon s(z) + \epsilon \right) p(z)= \left( 1 + \epsilon s(z) \right) p(z)
\end{equation}
Further, to prove that \(\tilde{\mathcal{P}}_\epsilon(Z) = \int_Z (1+\epsilon s(z))d\mathcal{P}(Z)\), we need the \textbf{Radon-Nikodym theorem}: 
\begin{theorem}[Radon-Nikodym Theorem]
Let \((\Omega, \mathcal{F}, \mathcal{P})\) be a probability space, and let \(\mathcal{\tilde{P}}_\epsilon\) be another probability measure on \((\Omega, \mathcal{F})\) such that measure \(\mathcal{\tilde{P}}_\epsilon\) is absolutely continuous with respect to measure \(\mathcal{P}\). Then there exists a \(\mathcal{P}\)-integrable function \(f: \Omega \to [0, \infty)\) such that for every \(A \in \mathcal{F}\),
\[
\mathcal{\tilde{P}}_\epsilon(A) = \int_A f \, d\mathcal{P}.
\]
The function \(f\) is called the Radon-Nikodym derivative and is often denoted by \(\frac{d\mathcal{\tilde{P}}_\epsilon}{d\mathcal{P}}\).
\end{theorem}

The theorem gives us a toolbox to "rescale" the measure \(\mathcal{\tilde{P}}_\epsilon\) with the measure \(\mathcal{P}\) and the \textbf{Radon-Nikodym derivative} function \(f\). Since Equation \ref{linearsf}, we the scale is indeed \(( 1 + \epsilon s(z) \), and thus, 
\begin{equation}\label{eq:measure}
\tilde{\mathcal{P}}_\epsilon(Z) = \int_Z \left( 1 + \epsilon s(z) \right) p(z)dz = \int_Z \left( 1 + \epsilon s(z) \right) d\mathcal{P}(Z).    
\end{equation}
\end{proof}

The definition and Equation \ref{linearsf} indeed reveal that the essence of the score function is a "direction pointer" (or compass). We could use the score function to specifically point the direction of the digress between the empirical estimator and the true statistical estimand. Suppose we have the estimand as \(\psi(\mathcal{P})\), the pathway derivative, or the gradient of the estimand in the direction of the score function\footnote{The gradient here is defined on the direction of the score function for the estimator. Indeed, the score function is also a gradient, but the gradient on the measurable set \(Z\): \(s(z) = \frac{\partial \log \tilde{p}_\epsilon(z)}{\partial \epsilon}\big|_{\epsilon = 0} = \nabla_z \log p(z) \).}, can be defined as: 
\begin{equation}
    \lim_{\epsilon \to 0}\frac{\psi(\tilde{\mathcal{P}}_\epsilon) - \psi(\mathcal{P})}{\epsilon} = \nabla_{s}\psi(\mathcal{P})
\end{equation}\\
Where we could regard \(\psi(\tilde{\mathcal{P}}_\epsilon)\) as a perturbed version of the true estimand \(\psi(\mathcal{P})\), adjusted by a small amount of \(\epsilon\) in the direction of some perturbation. \\

An obvious and important characteristic of the score function \(s(z)\) is that \(E[s(z)] = 0\) (so that \(\tilde{\mathcal{P}}_\epsilon(Z) = \int_Z (1+\epsilon s(z))d\mathcal{P}(Z)\) can be integrated to \(1\)) and \(\mathrm{Var}(s(z)) < \infty\)\footnote{In other words, score function \(s(z)\) is defined on the \(L_2^0\) space as its square-integral functions are integrable within finite values and the mean is zero. Besides the score function, the influence function, the efficient estimation, and the Cramer-Rao bound are all defined on the \(L^0_2\) space in our discussion.\label{l2space}}. Given this, we have an important corollary. Suppose we have a bivariate joint distribution \(\mathcal{P}(Y, X)\), clearly, we have the Bayesian rule \(\mathcal{P}(Y, X) = \mathcal{P}(Y \mid X)\mathcal{P}(X)\). We define the score function for \(\mathcal{P}(Y \mid X)\) and \(\mathcal{P}(X)\) separately as \(s_{Y |X}(x,y)\) and \(s_X(x)\). The score functions satisfy \(E[s_{Y |X}(x,y) \mid X] = 0\) and \(E[s_X(x)]=0\). Therefore, \(s_{Y |X}(x,y) \perp s_X(x)\). \\
\begin{proof}
    The target is to prove the expectation \(E[s_{Y |X}(x,y) s_X(x)] = 0.\) Rewrite the expectation with the law of conditional expectation:
    \begin{align*}
           E[s_{Y |X}(x,y) s_X(x)] & = E[E[s_{Y |X}(x,y) s_X(x) \mid X]] \\& = E[s_X(x) E[s_{Y |X}(x,y) \mid X]]  = E[s_X(x) \cdot 0] = 0. 
    \end{align*}
\end{proof}

With the orthogonal relationship between \(s_{Y |X}(x,y)\) and \(s_X(x)\), we could have \(s_{X,Y}(x,y) = s_{Y |X}(x,y) + s_X(x)\) and \(\nabla_{s_{X,Y}}\psi = \nabla_{s_{Y|X}}\psi + \nabla_{s_{X}}\psi \). Therefore, we could calculate the score function for the marginal distribution \(\mathcal{P}(X)\) and the conditional distribution \(\mathcal{P}(Y|X)\) within their own models and sum them together to get the score function for the joint distribution if the score function for the joint distribution is hard to capture. \\

\begin{proof}
    The joint distribution \(\mathcal{P}(Y, X)\) can be expressed using the chain rule of probability:
\[
p_{X,Y}(x,y) = p_{Y|X}(y|x) p_X(x)\iff \log p_{X,Y}(x,y) = \log p_{Y|X}(y|x) + \log p_X(x)
\]

Applying the definition of the score function, we differentiate both sides with respect to \(\epsilon\) and evaluate at \(\epsilon = 0\):
\[
s_{X,Y}(x,y) = \frac{\partial \log \tilde{p}_{X,Y, \epsilon}(x,y)}{\partial \epsilon} \bigg|_{\epsilon = 0} = \frac{\partial \log \tilde{p}_{Y|X, \epsilon}(y|x)}{\partial \epsilon} \bigg|_{\epsilon = 0} + \frac{\partial \log \tilde{p}_{X, \epsilon}(x)}{\partial \epsilon} \bigg|_{\epsilon = 0}
\]

Thus:
\begin{equation}\label{factorize}
s_{X,Y}(x,y) = s_{Y |X}(x,y) + s_X(x)
\end{equation}

By the definition of the gradient in the direction of the score function, we have:
\[
\nabla_{s_{X,Y}} \psi = \lim_{\epsilon \to 0} \frac{\psi(\tilde{\mathcal{P}}_\epsilon(X,Y)) - \psi(\mathcal{P}(X,Y))}{\epsilon}
\]

Since \(p_{X,Y}(x,y) = p_{Y|X}(x,y)p_{X}(x)\), we have:
\[
\nabla_{s_{X,Y}} \psi = \lim_{\epsilon \to 0} \frac{\psi(\tilde{\mathcal{P}}_\epsilon(X,Y) \tilde{\mathcal{P}}_\epsilon(X)) - \psi(\mathcal{P}(Y|X) \mathcal{P}(X))}{\epsilon}
\]

Given that \(\psi\) is influenced by the score functions \(s_{Y |X}\) and \(s_X\) independently, we have:
\[
\nabla_{s_{X,Y}} \psi = \lim_{\epsilon \to 0} \frac{\psi(\tilde{\mathcal{P}}_\epsilon(Y|X)) - \psi(\mathcal{P}(Y|X)}{\epsilon} + \lim_{\epsilon \to 0} \frac{\psi(\tilde{\mathcal{P}}_\epsilon(X)) - \psi(\mathcal{P}(X))}{\epsilon}
\]

Thus, we get:
\begin{equation}\label{sfgrad}
\nabla_{s_{X,Y}} \psi = \nabla_{s_{Y|X}} \psi + \nabla_{s_X} \psi    
\end{equation}
\end{proof}

In Equations \ref{factorize} and \ref{sfgrad}, we are actually \textbf{factorizing} the score function. An advantage of factorizing is that suppose the estimator is perturbed but only on one dimension after our appropriate factorization, then we can only calculate the change in the score function on that dimension and keep the others unchanged. We will use this advantage when we yield the efficient estimator later. \\

Indeed, we define the closure of the linear span of the score functions at \(\mathcal{P}(Z)\) the \textbf{tangent space} of \(\mathcal{P}(Z)\): \(\mathcal{T}(\mathcal{P}(Z))\). As we showed in the factorization steps, we could use the set of the linear combination of the score functions to define the tangent space\footnote{Since the score functions are defined on the \(L_2^0\) space, we could define the tangent space as the set of the square-integrable functions with respect to \(\mathcal{P}(Z)\) whose means are \(0\): \[
\mathcal{T}(\mathcal{P}(Z)) = \left\{ h \in L_2(\mathcal{P}(Z)) : E_{\mathcal{P}(Z)}h(x) = 0 \right \}.
\] \label{tangent2} }
:
\[
\mathcal{T}(\mathcal{P}_Z) = \left\{ h : h(z) = \sum_{i} \alpha_i s_{\epsilon}(z), \quad \alpha_i \in \mathbb{R}, \quad \epsilon \in [0,1] \right\} 
\]\\

Since any score function belonging to the tangent space of the specific measure can be used during factorization, therefore, we are actually factorizing the tangent space. In the above bivariate example, we could write the factorization as: \(\mathcal{T}_{X,Y} = \mathcal{T}_{Y|X} \oplus \mathcal{T}_X\), where \(\mathcal{T}_{Y|X}\) is defined as the tangent space associated with the conditional distribution \(\mathcal{P}(Y|X)\): \(\mathcal{T}_{Y|X} =  \left\{ h_{Y|X}(x,y): E[h_{Y|X}(x,y) \mid X = x] = 0 \text{ for all } x \right\}\). Correspondingly, \(\mathcal{T}(X)\) denotes the tangent space associated with the marginal distribution \(\mathcal{P}(X)\): \(\mathcal{T}(X) = \left\{ h_X(x) : E[h_X(X)] = 0  \text{ for all } x \right\}\). The symbol \(\oplus\)  denotes the direct sum, indicating that these spaces are orthogonal.\\

If the tangent space contains all the square-integrable and mean-zero functions (the same as the \(L_2^0\) space, see Footnotes \ref{l2space} and \ref{tangent2}), the model which specifies the measure \(\mathcal{P}(Z): \mathcal{M}: \mathcal{P}(Z) \in \mathcal{M} \) is a \textbf{saturated model}. For saturated models, the tangent space exists for all directions, meaning that if we move towards any direction, we are still in the model. Suppose our models are fully nonparametric; then, our model is saturated since no restrictions impede us. Otherwise, the model is not saturated. For instance, a general causal model is saturated, but an RCT causal model is not because the propensity in the randomized trial is fixed.  \\

Finally, we could use the idea in the score function to define regularity. If we define \(\tilde{\mathcal{P}}_n := \tilde{\mathcal{P}}_{\epsilon = 1/\sqrt{n}}\). Suppose \(\tilde{\mathbb{P}}_n\) is the empirical distribution of \(n\) samples drawn from the perturbed distribution \(\tilde{\mathcal{P}}_n\). According to Equation \ref{regu}, regularity is defined as \(n\) increases, divergence between the empirical estimator \(\hat{\psi}(\tilde{\mathbb{P}}_n)\) and the statistical estimand \(\psi(\tilde{\mathcal{P}}_n)\) converges in a fixed distribution. In this setting, with \(n\) increases, \(\epsilon\) shrinks, and \(\tilde{\mathcal{P}}_\epsilon\) gets closer to \(\mathcal{P}\), and the estimates are increasingly accurate and reflective of the true distribution as the sample size grows and the perturbation diminishes. 
\subsection{Asymptotical Linearity and Influence Function}
Contradicting to the property of regularity, asymptotical linearity is more intuitive. Multidimensional CLT has suggested that, the divergence between the sample mean of a series of IID random vectors (\(\bar{X}_n = \frac{1}{n}\sum_{k=1}^n X_k\)) and the expectation (\(\mu = E[X]\)) converges to the normal distribution with mean \(0\) and variance \(\Sigma = E[(X_i - \mu)(X_i-\mu)^T]\)\footnote{Proving the multidimensional CLT requires prior knowledge on characteristic functions, Levy's Continuity Theorem, and Cramer-Wold device. The characteristic function of the real-valued random variable defines the probability distribution-- in other words, two distinct distributions with the same characteristic function are identically the same distribution. Levy's continuity theorem states that if the characteristic functions of a sequence of random variables pointwise converge towards the characteristic function of a limiting random variable, then the sequence of random variables converges in distribution towards the limiting random variable. Cramer-Wold device states that a sequence of random variables converging to a limiting random variable is equivalent to the scalar of the sequence of random variables converging to the scalar of the limiting random variable. With these backgrounds, we have the proof: 

      We first define the normalized sample mean vector: given that \(\bar{X}_n = \frac{1}{n} \sum_{i=1}^{n} X_i\), where \(X_i\) are i.i.d. random variables with mean \(\mu = E[X_i]\) and covariance matrix \(\Sigma = E[(X_i - \mu)(X_i-\mu)^T]\), let \(Z_n = \sqrt{n}(\bar{X}_n - \mu )\). Then we find the characteristic function of \(Z_n\):
    \[
\varphi_{Z_n}(t) = E\left[ e^{i t^T Z_n} \right] = E\left[ e^{i t^T \sqrt{n} (\bar{X}_n - \mu)} \right] = E\left[ e^{i t^T \sqrt{n} \left( \frac{1}{n} \sum_{i=1}^{n} X_i - \mu \right)} \right]
\]
Using the Taylor expansion of the exponential function \(e^{ix} \approx 1 + ix - \frac{x^2}{2}\), we get:
\[
E\left[ e^{i t^T \frac{X_i-\mu}{\sqrt{n}}} \right] \approx 1 + i t^T \frac{E[Y_i]}{\sqrt{n}} - \frac{1}{2} t^T \frac{E[(X_i-\mu) (X_i-\mu)^T]}{n} t
\]
Since \(E[X_i-\mu] = 0\) and \(E[(X_i-\mu)(X_i-\mu)^T] = \Sigma \), this simplifies to:
\[
E\left[ e^{i t^T \frac{X_i-\mu}{\sqrt{n}}} \right] \approx 1 - \frac{1}{2} t^T \frac{\Sigma}{n} t
\]\\

Therefore,
\[\varphi_{Z_n}(t) \approx \left( 1 - \frac{1}{2} t^T \frac{\Sigma}{n} t \right)^n \to e^{- \frac{1}{2} t^T \Sigma t} \text{ as }n \to \infty
\]\\

By Levy's Continuity Theorem, this implies that \(Z_n\) converges in distribution to a multivariate normal distribution:
\[
Z_n \xrightarrow{d} \mathcal{N}(0, \Sigma)
\]
Therefore, we have proved the multidimensional Central Limit Theorem:
\[
\sqrt{n}(\bar{X}_n - \mu) \xrightarrow{d} \mathcal{N}(0, \Sigma)
\]  
For the specific details, see \citeauthor{vandervaart1998} \citeyear{vandervaart1998}: Ch.2, pp.12-16. }. 
\begin{equation}\label{mltclt}
    \sqrt{n}(\bar{X}_n - \mu) \xrightarrow{d} N(0, \Sigma)
\end{equation}
Notice the similarities between Equations \ref{AL2} and \ref{mltclt}, the relationship between asymptotical linearity and asymptotical normality can't be more clear: in fact, if an estimator is asymptotically linear, then the CLT can be applied to the linear part of the estimator. We rewrite Equation \ref{AL} into a linear form:
\begin{equation}\label{AL4}
    \psi(\mathbb{P}_n) = \psi(\tilde{\mathcal{P}}) + \frac{1}{n}\sum_{i=1}^n \phi(\psi; \mathbb{P}_n; Z_i) + o_p(n^{-1/2})
\end{equation}
The term \(\frac{1}{n}\sum_{i=1}^n \phi(\psi; \mathbb{P}_n; Z_i) \) can be treated as a sum of i.i.d random variables (for instance, if the estimator is the sample mean \(\psi(\mathbb{P}_n) = \bar{X}_n\), the term \(\phi(\psi; \mathbb{P}_n; X_i) = (X_i - \mu)\) so that we have Equation \ref{AL4} written as \(\bar{X}_n = \mu + \frac{1}{n}\sum_{i=1}^n(X_i - \mu)\)). As we noted in Equation \ref{AL2}, the variance of the term is the covariance for the converged normal distribution. In asymptotical linearity, the term is called \textbf{influence function}, as it shows the sensitivity of the estimator to small changes or perturbations in the data. It is named the influence function as it measures how each data point in the dataset (\(Z_i \in Z\)) influences the overall estimate. If the estimator is unbiased (which for all the RAL estimators are true), the influence function should have a zero mean and finite variance (defined on the \(L^0_2\) space, see Footnote \ref{l2space}).\\

Secondly, we notice that the convergence rate in both the CLT and the asymptotical linearity is \(1/\sqrt{n}\) (as we introduced before, the term \(o_p(n^{-1/2}\) is called the second-order remainder). The convergence rate in the CLT is understandable: the expectation and variance of the sum of the \(n\) random variables \(S_n = \sum_{i=1}^n X_i\) are separately \(E[S_n] = n\mu\) and \(\text{Var}(S_n) = n\sigma^2\). Therefore, during standardization, we have: \(\frac{S_n - E[S_n]}{\sqrt{\text{Var}(S_n)}} = \frac{S_n-n\mu}{\sqrt{n\sigma^2}}\). Since the standardization involves dividing by \(\sqrt{n\sigma^2}\), we have the convergence rate \(n^{-1/2}\) to balance the increase in total variability as the sample size increases. Similarly, in asymptotically linear settings, since \(\frac{1}{n}\sum_{i=1}^n \phi(\psi; \mathbb{P}_n; Z_i) \) involves averaging the influence function over the sample, which is expected to converge to the normal distribution of mean and variance of \(\phi(\psi; \mathbb{P}_n; Z_i)\), and to balance the decrease in the asymptotical variance as the sample size \(n\) increases, the convergence rate \(n^{-1/2}\) emerges naturally like we elaborated on the CLT case. \\

The influence function and the second-order remainder are the core parts of asymptotic analysis, and analogizing them with the CLT expression is only one perspective. The expression \(\phi(\psi; \mathbb{P}_n; Z_i)\) can not only be regarded as the influence function, it is also the pathway derivative, the gradient, and the Neyman orthogonal score. \\

Consider the core question of asymptotical analysis. Our target is to find the unbiased estimator \(\psi(\mathbb{P}_n)\) for the statistical estimand \(\psi(\mathcal{P})\) with the least statistical variance. Consider we have a continuous, smooth path defined as \(\{\tilde{\mathcal{P}}_\epsilon\}\) starting from \(\mathcal{P}(Z) = \mathcal{P}\) where \(\epsilon = 0\) ending at \(\tilde{\mathcal{P}}(Z) = \tilde{\mathcal{P}}\) where \(\epsilon = 1\). We suppose \(\psi(\mathcal{P})\) is the estimand (\(\psi(\mathcal{P})\)), \(\psi(\tilde{\mathcal{P}})\) is our estimator ( \(\psi(\mathbb{P}_n)\)). In the last session, we have shown that we could go from \(\psi(\tilde{\mathcal{P}})\) to approach \(\psi(\mathcal{P})\) by making \(\epsilon \to 0\) (as \(n \to \infty\)). Now, we consider the static "snapshot," which allows us to use \(\psi(\mathcal{P})\) to estimate \(\psi(\tilde{\mathcal{P}})\). An intuitive idea here is to construct the \textbf{functional (distributional) Taylor expansion}: 
\begin{lem}
    Distributional Taylor Expansion: For a function \(f(x)\) which is differentiable in \([X_0, X_1]\), the distributional Taylor expansion can be expressed as:
    \begin{equation}\label{taylor}
        f(X_0) \approx f(X_1) + \nabla f(X_1)(X_0 - X_1) + \frac{1}{2}(X_0 - X_1)\nabla^2f(X_1)(X_0 - X_1)^T...
    \end{equation}
    or,
    \[
    f(X_0) \approx f(X_1) + \frac{\partial f(x)}{\partial x}\big|_{x = X_1}(X_0 - X_1) + \frac{1}{2}\frac{\partial^2f(x)}{\partial x^2}\big|_{x = X_1}(X_0 - X_1)^2...
    \]
\end{lem}

We can apply Equation \ref{taylor} in our analysis. For the function \(\psi(\tilde{\mathcal{P}}_\epsilon)\) in the domain \(\epsilon \in [0,1]\), we have\footnote{This is also called \textbf{"von Mier Expansion"} in some literature. }:
\begin{equation}\label{eq:taylordecomp}
    \psi(\mathcal{P}) \approx  \underbrace{\psi(\tilde{\mathcal{P}})}_{\text{naive plug-in estimator}}  +\underbrace{\frac{\partial \psi(\tilde{\mathcal{P}}_\epsilon)}{\partial \epsilon}\bigg|_{\epsilon = 1}(0 - 1)}_{\text{first-order bias correction}} + \underbrace{\frac{1}{2} \frac{\partial^2 \psi(\tilde{\mathcal{P}}_\epsilon)}{\partial \epsilon^2}\bigg|_{\epsilon = 1}(0 - 1)^2}_{\text{second-order remainder}} + \cdots
\end{equation}

\begin{figure}[ht]
    \centering
    \includegraphics[width=\textwidth]{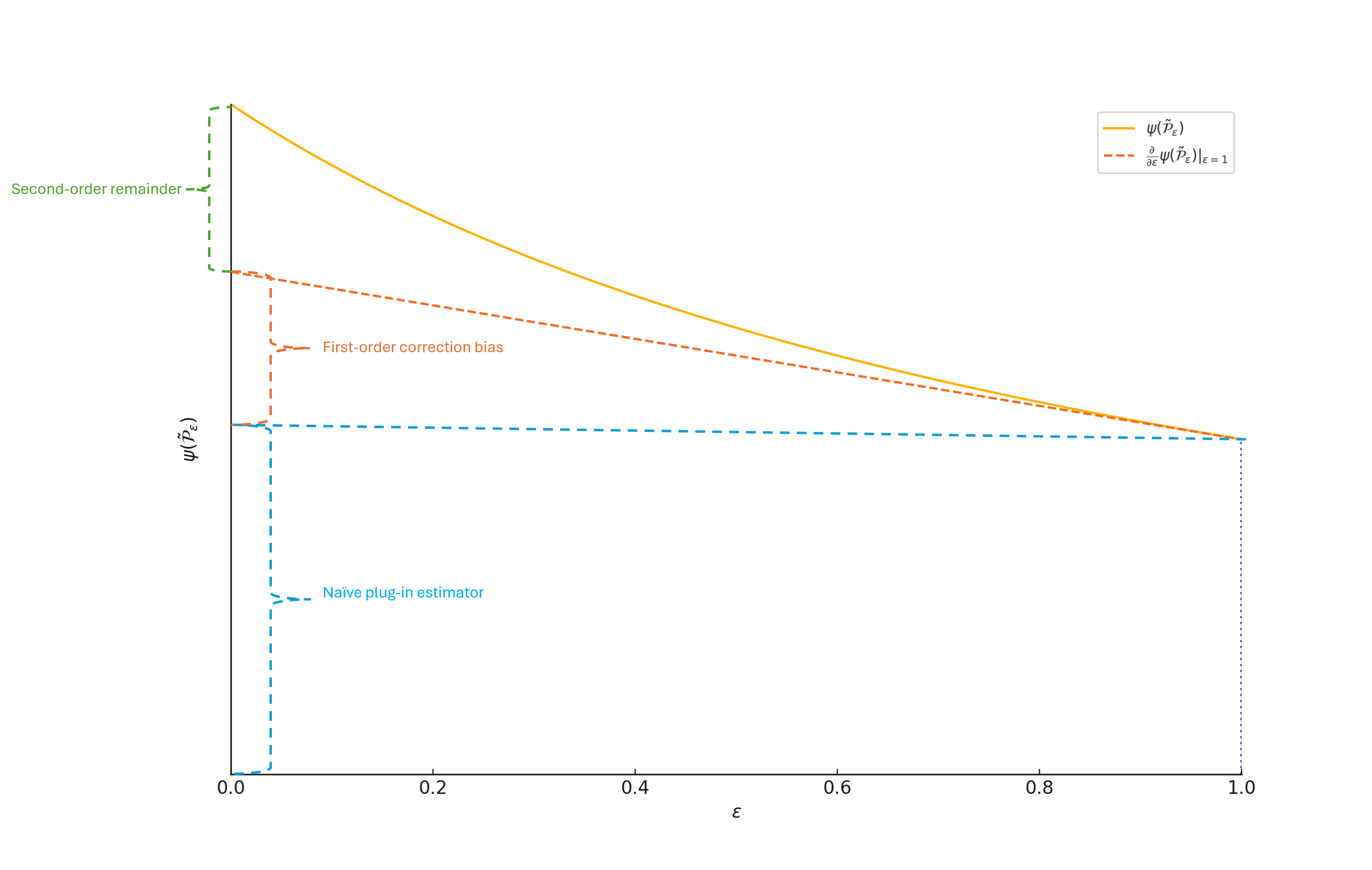}
    \caption{Illustration on Distributional Taylor Expansion}
    \label{fig:taylor}
        \raggedright \small \textbf{Note:} This is a simplified illustration of the Distributional Taylor expansion. Indeed, \(\tilde{\mathcal{P}}\) and \(\mathcal{P}\) are two measures, but we just simplify them as two values. If so, the direction of the score function should be horizontal, and the direction of the gradient of the score function on the estimator should be the tangent line of \(\psi\) at \(\psi(\mathcal{P})\), which has the same direction as the influence function. Thus, the angle of score and influence function points to the same direction as the gradient. 
\end{figure}

Therefore, 
\[
\psi(\tilde{\mathcal{P}}) - \psi(\mathcal{P}) \approx \underbrace{\frac{\partial \psi(\tilde{\mathcal{P}}_\epsilon)}{\partial \epsilon}\bigg|_{\epsilon = 1}}_{\text{first-order bias correction}} - 
 \underbrace{\frac{1}{2} \frac{\partial^2 \psi(\tilde{\mathcal{P}}_\epsilon)}{\partial \epsilon^2}\bigg|_{\epsilon = 1}}_{\text{second-order remainder}}
\]
To make the above decomposition more intuitive, we illustrate the ideas in Figure \ref{fig:taylor}, which shows how to use results of \(\psi(\mathcal{\tilde{P}})\) to approach \(\psi(\mathcal{P})\). We will discuss the details as follows. 
\subsubsection{First-order bias correction and influence functions}
We first analyze the first-order bias correction term. Due to the chain rule, we have:
\begin{equation}\label{chain}
    \frac{\partial }{\partial \epsilon}\psi(\tilde{\mathcal{P}}_\epsilon)\bigg|_{\epsilon = 1} = \frac{\partial \psi(\tilde{\mathcal{P}}_\epsilon)}{\partial \tilde{\mathcal{P}}_\epsilon}\cdot\frac{\partial \tilde{\mathcal{P}}_\epsilon}{\partial \epsilon}\big|_{\epsilon = 1} = \sum_{i=1}^n \frac{\partial \psi(\tilde{\mathcal{P}}_\epsilon)}{\partial \tilde{\mathcal{P}}_\epsilon(Z_i)}\cdot\frac{\partial \tilde{\mathcal{P}}_\epsilon(Z_i)}{\partial \epsilon}\big|_{\epsilon = 1}  = \sum_{i=1}^n \frac{\partial \psi(\tilde{\mathcal{P}}_\epsilon)}{\partial p_\epsilon(Z_i)}\cdot (p_1(Z_i) - p_0(Z_i))
\end{equation}
as \(p_\epsilon(Z_i)\) refers to the probability mass function for measure \(\mathcal{P}\) at the specific point \(Z_i\). The equation shows how each probability mass may affect the change in the estimator \(\psi(\tilde{\mathcal{P}}_\epsilon)\). We may recall that we introduced the influence function because we would like to reveal how a slight perturbation at any data point in the dataset \(Z\) affects the estimator \(\psi(\tilde{\mathcal{P}}_\epsilon)\). In order to understand the relationship between \(\frac{\partial}{\partial \epsilon}\psi(\tilde{\mathcal{P}}_\epsilon)\) and the influence function \(\phi(\psi; \mathbb{P}_n, Z_i)\), we introduce the concept of \textbf{Gateaux derivative}:\\

\begin{lem}
Definition of the influence function with Gateaux derivative: The Gateaux derivative of \(\psi\) at \(\mathcal{P}\) in the direction of another distribution G is defined as:
\begin{equation}\label{gateaux}
\mathbf{D}_\psi(\mathcal{P}; \mathcal{G}) = \lim_{\epsilon \to 0} \frac{\psi((1-\epsilon)\mathcal{P} + \epsilon \mathcal{G}) - \psi(\mathcal{P})}{\epsilon}    
\end{equation}

Now suppose for distribution \(\mathcal{P}\) we have a small perturbation at a point \(z\), we write the perturbed distribution as:
\[
\tilde{\mathcal{P}}_\epsilon = \mathcal{P} + \epsilon(\delta_z - \mathcal{P}) \Rightarrow \psi(\tilde{\mathcal{P}}_\epsilon) = \psi(\mathcal{P} + \epsilon(\delta_z - \mathcal{P}))
\]
Where \(\delta_z\) stands for the point mass distribution at \(z\) \footnote{Generally, we call \(\delta_z\) the \textbf{Dirac delta function} at \(z\). However, it is not a traditional function; rather, it is a distribution \(\delta_a\) which is concentrated at a single point of \(z\). In some literature, introducing the Dirac delta function is a trick for "point mass contamination", which introduces a particular kind of path where the destination is a distribution that places all of its mass at the point \(z\).}. Now that the Gateaux derivative of \(\psi\) at \(\mathcal{P}\) in the direction of \(\delta_z \) is:
\begin{equation}\label{derivative}
    \mathbf{D}_\psi(\mathcal{P}; \delta_z ) = \lim_{\epsilon \to 0} \frac{\psi(\mathcal{P} + \epsilon(\delta_z - \mathcal{P})) -\psi(\mathcal{P}))}{\epsilon} 
\end{equation}
\end{lem}
Gateaux derivative expressed in Equation \ref{derivative} is the mathematical definition of the influence function of \(\phi(\psi; \mathcal{P})\), as it describes the impact of a small change in the data on the given functional \(\psi\). Since the influence function describes an infinitesimal perturbation in the distribution (measure) \(\mathcal{\tilde{P}}_\epsilon\) at point \(z\), we could regard the influence function as a dichotomous representation of the functional derivative \(\frac{\delta \psi(\mathcal{\tilde{P}}_\epsilon)}{\delta \tilde{p}_\epsilon(z)}\), or the functional derivative as the influence function in the context of continuous distributions. We could have some mathematical transformation with the definition to reveal its relationship with the first-order bias. We first use partial differentiation to express the equation:
\[
\phi(\psi; \tilde{\mathcal{P}}; z) = \lim_{\epsilon \to 0} \frac{\psi(\tilde{\mathcal{P}} + \epsilon(\delta_z - \tilde{\mathcal{P}})) -\psi(\tilde{\mathcal{P}}))}{\epsilon}
\]
Integral the influence function on the direction of \(\mathcal{P} - \tilde{\mathcal{P}}\), we have:
\begin{equation}\label{integral}
\frac{\partial \psi\big(\tilde{\mathcal{P}} + \epsilon(\mathcal{P} - \tilde{\mathcal{P}})\big)}{\partial \epsilon} \bigg|_{\epsilon = 0}
 = \int \phi(\psi; \tilde{\mathcal{P}}; z)(p(z) - \tilde{p}(z)) dz    
\end{equation}

 Notice on the left side, the numerator is actually \(\tilde{\mathcal{P}} + \epsilon(\mathcal{P} - \tilde{\mathcal{P}}) = \tilde{\mathcal{P}}_\epsilon\),  the right side of Equation \ref{chain}, which has a similar structure. Therefore,
 \begin{align*}
      \frac{\partial }{\partial \epsilon}\psi(\tilde{\mathcal{P}}_\epsilon)\bigg|_{\epsilon = 1} &= - \int \phi(\psi; \tilde{\mathcal{P}}; z)(p(z) - \tilde{p}(z)) dz = - \int\phi(\psi; \tilde{\mathcal{P}}; z)p(z) dz \\ &= -\frac{1}{n}\sum_{i = 1}^n \phi(\psi; \tilde{\mathcal{P}}; Z_i)
 \end{align*}

 Because \(\int \phi(\psi; \tilde{\mathcal{P}}; z)\tilde{p}(z) dz = 0\)\footnote{Similarly, we could also have the derivation: \(\frac{\partial}{\partial \epsilon}\psi(\tilde{\mathcal{P}}_\epsilon)\bigg|_{\epsilon = 0} = \frac{1}{n}\sum_{i=1}^n \phi(\psi; \mathcal{P}, Z_i)\). Notice the left side of the equation, \(\frac{\partial}{\partial \epsilon} \psi(\tilde{\mathcal{P}}_\epsilon)\bigg|_{\epsilon = 0}\), is indeed the gradient on the direction of the score function on the estimator \(\nabla_s \psi(\mathcal{P})\).}. We are very familiar with the right side of the equation: this is the expectation of the influence function. This is to say, the influence function can be regarded as a pathwise derivative, and we could approximate \(\psi(\mathcal{P})\) with the naive plug-in estimator \(\psi(\mathbb{P}_n)\) plus the expectation of the influence function for \(\mathbb{P}_n\)). We define this estimator as the "one-step" estimator \citep{Fisher2021}:
 \begin{equation}\label{1-step}
     \hat{\psi}(\mathbb{P}_n)_{\text{1-step}} \approx \psi(\mathbb{P}_n) + \sum_{i = 1}^n\phi(\psi; \mathbb{P}_n; Z_i)
 \end{equation} 
 \\
 
 The Gateaux derivative also reveals the relationship between the gradient to the score function and the influence function. Recall our definition of the gradient in the direction of the score function on the estimator, we have:
 \[
 \nabla_s \psi(\mathcal{P}) = \lim_{\epsilon \to 0} \frac{\psi(\tilde{\mathcal{P}}_\epsilon) - \psi(\mathcal{P})}{\epsilon} = \frac{\partial}{\partial \epsilon}\psi(\tilde{\mathcal{P}}_\epsilon)\bigg|_{\epsilon = 0}
 \]

Similarly, using distributional Taylor expansion, we have:
\[
\psi(\tilde{\mathcal{P}}_\epsilon)\bigg|_{\epsilon = 0} \approx  \psi(\mathcal{P}) + \epsilon \left. \frac{\partial \psi}{\partial \epsilon} \right|_{\epsilon=0} + O(\epsilon)
\]
And use the chain rule to analyze the first-order term :

\[
\left. \frac{\partial\psi}{\partial\epsilon} \right|_{\epsilon=0} = \int \frac{\delta \psi}{\delta p(z)} \left. \frac{\partial \tilde{p}_\epsilon(z)}{\partial \epsilon} \right|_{\epsilon=0} dz.
\]

Based on Equation \ref{linearsf} describing the relationship between \(p(z)\) and \(\tilde{p}_\epsilon(z)\):\(\tilde{p}_\epsilon(z) = \left( 1 + \epsilon s(z) \right) p(z)\), we have:
\[
\left. \frac{\partial \tilde{p}_\epsilon(z)}{\partial \epsilon} \right|_{\epsilon=0} = \frac{\partial}{\partial\epsilon}\left( 1 + \epsilon s(z) \right) p(z)\bigg|_{\epsilon = 0}= s(z) p(z).
\]
Therefore, 
\[
\left. \frac{d\psi}{d\epsilon} \right|_{\epsilon=0} = \int \frac{\delta \psi}{\delta p(z)} s(z) p(z) \, dz.
\]
Note the right side of the expression can be expressed as the form of expectation under the original distribution \(p(z)\):
\[
\int \frac{\delta \psi}{\delta p(z)} s(z) p(z) \, dz = \mathbb{E}_{\mathcal{P}} \left[ \frac{\delta \psi}{\delta p(z)} s(z) \right].
\]

As we mentioned, the form \(\frac{\delta \psi(P)}{\delta p(z)}\) can be regarded as the continuous form of the influence function \(\phi(\psi; \mathcal{P}; z)\)\footnote{In some literature, the definition of the influence function with the expectation expression is the score-based definition of the influence function, and the score function can also be regarded as \(s_0(z) = \frac{\partial}{\partial \epsilon}\log [p(z) +\epsilon(\tilde{p}_\epsilon (z) - p(z))]\big|_{\epsilon = 0} =\frac{\tilde{p}_\epsilon(z)-p(z)}{p(z)}\), which is the same as the definition in Equation \ref{score2}.}. Therefore, we finally obtain:
\begin{equation} \label{central}
   \nabla_s \psi(\mathcal{P}) = \mathbb{E}_{\mathcal{P}} \left[\phi(\psi; \mathcal{P}; z) s(z) \right].
\end{equation}

Equation \ref{central} is called the \textbf{central identity for influence functions}. This equation reveals that the gradient of the functional \(\psi\) with respect to the score function \(s(z)\) of the estimator is given by the expectation of the product of the influence function \(\phi(\psi; \mathcal{P}; z)\) and the score function \(s(z)\) under the distribution \(\mathcal{P}\). This relationship indicates that the direction of the gradient is aligned with the interaction between the influence function and the score function (as the expectation of the score and the influence functions under RAL are both zero). This equation is the most useful tool for extracting the influence function or validating if the influence function is correct, especially for efficient influence functions for the saturated models. We will give a specific example of the usage in the next subsection.\\

Also, since the influence functions \(\Phi\) are indeed a gradient of the estimator \(\psi\), we may have the "gradient algebra" rules for the influence functions. For instance, the chain rule: \(\phi( g(\psi)) = g'(\psi)\phi(\psi)\); and the product rule: \(\phi(\psi_1\psi_2) = \phi(\psi_1)\psi_2 + \psi_1\phi(\psi_2)\). Similar rules are also applied to the score functions and the gradient on the score functions.\\

Finally, the influence function can be regarded as the \textbf{Neyman's orthogonal score}. Originally, Neyman orthogonality refers to a score function that yields robust estimations for small perturbations. Neyman's orthogonal score is defined as a function \(f(Z; \hat{\psi};\hat{\epsilon})\) (\(Z\) denotes the measurable set, \(\hat{\psi}\) denotes the estimator of interest, and \(\hat{\epsilon}\) is the perturbation) which is orthogonal to the perturbation when we are at the true (population) value \citep{chernozhukov2018a}: \[\frac{\partial}{\partial \epsilon}E[f(Z; \hat{\psi}; \hat{\epsilon})\big|_{\hat{\epsilon} = 0} ] = 0\].\\

The influence function \footnote{The population true values in general regarded as \(\epsilon_0\), and therefore, the condition is \(\hat{\epsilon} = \epsilon_0\).} is intrinsically equivalent to Neyman's orthogonal score simply because it is the mean-zero score function which measures the gradient on the estimator towards the perturbation. Recall we have defined the gradient of an estimator on the direction of the score function (Equation \ref{sfgrad}): \(\nabla_{s}\psi(\mathcal{P}) = \lim_{\epsilon \to 0}\frac{\psi(\tilde{\mathcal{P}}_\epsilon) - \psi(\mathcal{P})}{\epsilon}\). Now suppose \(s_z\) denotes the score function of the perturbed distribution \(\delta_z\) and \(\tilde{\mathcal{P}}_\epsilon = (1-\epsilon)\mathcal{P} + \epsilon\delta_z\). Thus, we have:
\begin{equation}
    \nabla_{s_z} \psi(\mathcal{P}) = \lim_{\epsilon \to 0}\frac{\psi(\mathcal{P}+\epsilon(\delta_z - \mathcal{P}) - \psi(\mathcal{P})}{\epsilon} = \frac{\partial}{\partial\epsilon}[\psi(\mathcal{P}+\epsilon(\delta_z - \mathcal{P})]\big|_{\epsilon = 0} = \phi(\psi; \mathcal{P}; z)
\end{equation}

Obviously, the right side of the equation is the Gateaux derivative definition of the influence function. Moreover, due to the central identity of the influence function, we know the gradient in the direction of \(s_z\) should be equivalent to the covariance of the corresponding score function and the influence function. Thus, we get:
\[
\nabla_{s_z} \psi(\mathcal{P}) = \phi(\psi; \mathcal{P}; z) = E[\phi(\psi; \mathcal{P}; z) s_z(Z)]
\]
We bring up the perspective of Neyman's orthogonality on the influence function since most of the econometric literature on \textbf{doubly/debiased machine learning} starts with Neyman's orthogonality to find the robust estimator of the parameter of interests (for instance, the estimator on the ATE), which is the basis of the doubly/debiased machine learning method(\citeauthor{chernozhukov2018a} \citeyear{chernozhukov2018a}a; \citeauthor{che2018} \citeyear{che2018}b). Indeed, we illustrate here that the robust/debiased estimator based on the Neyman orthogonality is mathematically intrinsic to the efficient estimator based on the influence function. As for the double robustness, they apply the sample-splitting method to ensure the asymptotical validity of the convergence process, which we will discuss below. 

\subsubsection{Second-order remainder}
Once we develop the one-step estimator from the first-order bias correction term, we would like to evaluate the quality of the estimator-- whether it is a "good" proxy of the true estimand with observational data on hand. Generally, the quality of the estimator is decided by several factors: the sample size (which determines the effect of bias correction since we sum up the influence functions), the smoothness of the functional form (to be discussed below), and the choice of the measure \(\tilde{\mathcal{P}}\) (when \(\epsilon = 1\)). \\

To introduce the idea of \textbf{smoothness} of the functional form, we turn our focus to the second-order remainder, which is the gap between the one-step estimator and the true estimand. We denote it as \(R_2(\mathbb{P}_n, \mathcal{P}) = \psi(\mathcal{P}) - \psi(\mathcal{P})_{\text{one-step}}\). This part is quite similar to the \textbf{loss function} in machine learning terms, as it measures the difference between the predictor (estimator) \(\psi(\mathcal{P}_n)\)and the true value (estimand) \(\psi(\mathcal{P})\). \\

In Figure \ref{fig:taylor}, we simply project the continuous measures of \(\tilde{\mathcal{P}}_\epsilon\) on the x-axis and take two points on the axis separately as \(\tilde{\mathcal{P}}\) and \(\mathcal{P}\). Therefore, the direction of perturbation parallels to the x-axis. We may also imagine that starting from \(\mathcal{P}\), we may have infinite choices of directions (spanning the tangent space), and the hyperplane \(\psi\) intersects with each dimension along \(\epsilon\). So, we have estimators \(\psi(\tilde{\mathcal{P}})\) in all dimensions. If we would like to compare the smoothness of the estimators stretching from different dimensions over different distances, we need a process to "standardize" the measures onto one dimension. Therefore, we come up with a standardized measure \(\mathbf{P}\):
\[
\mathbf{P} := \mathcal{P} + \frac{\Delta}{||\mathcal{\tilde{P}} - \mathcal{P}||_2}(\mathcal{\tilde{P}} - \mathcal{P}) 
\]
where \(||\mathcal{\tilde{P}} - \mathcal{P}||_2\) is the \(L_2\) distance between the two measures \(\mathcal{\tilde{P}}\) and \(\mathcal{P}\): \(||\mathcal{\tilde{P}} - \mathcal{P}||_2 = \sqrt{\int (\tilde{p}(z) - p(z))^2 dz}\). As we can tell from the definition, \(\Delta\) shows the absolute distance from the standardized measure to the original measure: \(\Delta = ||\mathbf{P} - \mathcal{P}||_2\). The standardization process makes it possible to compare and analyze the smoothness of the estimators uniformly across different dimensions and distances shown below.\\

With the definition of the standardized measure, we could define the \(j\)-th order smoothness as the \(j\)-th order derivative of the estimator \(\psi(\mathbf{P}\) (if it is \(j\)-th order differentiable) with respect to the absolute distance \(\Delta\) converges to a constant number if \(\Delta\) approaches zero: \[
\frac{\partial^j}{\partial\Delta^j}\psi(\mathbf{P})\big|_{\Delta \to 0} = O(1). 
\] 
Fisher and Kennedy (\citeyear{Fisher2021}) have a very accurate metaphor to help understand the idea of "smoothness": it is like the "magician tablecloth trick." If the estimators are more smooth, changing the parameter \(\Delta\) (like pulling the tablecloth) will have less impact on the estimator. Since many (nonparametric) machine learning models use gradient-based methods for optimization, the smooth differentiable property of the estimator (and the convexity of the function) ensures the model converges to the global minimum, and facilitates the gradient optimization. \\

Specifically for the second-order smoothness, it allows the functional form of the estimator \(\psi\) to change nonlinearly with the perturbation, which is to say, the second derivative of \(\psi\) with respect to \(\Delta\) exists and is bounded. If the second-order smoothness is allowed, usually researchers presume that the linear approximations are insufficient to capture the true estimand. However, adding the term will help improve the error bound of our estimator. Accounting for the curvature of the function:

\[
\left| \psi(\tilde{P}_\epsilon) - \psi(P) - \psi'(P)(\tilde{P}_\epsilon - P) - \frac{1}{2} \psi''(P)(\tilde{P}_\epsilon - P)^2 \right| = O(\epsilon^2)
\]

the \(O(\epsilon^2)\) term indicates that the error decreases quadratically with the size of the perturbation, providing a much tighter bound compared to the linear \(O(\epsilon)\) term from the first-order approximation:
\[
\left| \psi(\tilde{P}_\epsilon) - \psi(P) - \psi'(P)(\tilde{P}_\epsilon - P) \right| = O(\epsilon)
\]

However, in the efficient estimation we are discussing in this thesis, we assume that the second-order remainder term becomes asymptotically negligible faster than the rate of \(1/\sqrt{n}\). In other words, as the sample size \(n\) increases, \(R_2\) diminishes faster than \(1/\sqrt{n}\) and makes its impact on the overall estimation asymptotically insignificant:
\begin{ass}[Negligiblity of the Second-Order Remainder]\label{ass:second}
For the second-order remainder \(R_2\), we have the assumption that:
\[
R_2 = o_p\left(\frac{1}{\sqrt{n}}\right).
\]

Which indicates:
\[
\forall \epsilon > 0, \quad \forall c > 0, \quad \Pr\left(\left|R_2\right| > \frac{c}{\sqrt{n}}\right) \to 0 \quad \text{as} \quad n \to \infty.
\]    
\end{ass}
If the Assumption is violated, which suggests that the \(R_2\) term converges no faster than the rate of \(1/\sqrt{n}\), meaning that the \(R_2\) term cannot be regarded as insignificant as the sample size grows (under large sample size), so the first-step estimator may not be asymptotically valid to be the approximation on the true estimand. Under such a scenario, sensitivity analyses or higher-order correction terms are needed (for instance, adding the second-order smoothness term as we elaborated above). Luckily, many machine learning models, including but not restricted to neural networks, highly adaptive lasso, gradient boosting, nearest neighbors, and regressive methods, have the property of consistency in the prediction, and we do not need to worry about the convergence rate for the second-order remainder applying these models\footnote{The property is called \textbf{\(L_2\) consistency}, which means that as the sample size increases, prediction error of the model (measured in terms of \(L_2\) norm or simply MSE) converges to the lowest possible prediction error for the underlying data distribution.}.

\subsubsection{Empirical Process Convergence}
The second-order remainder term is \(1/\sqrt{n}\)-negligible suggests that we would like the empirical data distribution, as the sample size increases, to converge to the true distribution at the pace faster than \(1/\sqrt{n}\) for all other parts than the one-step estimator. However, in Equation \ref{1-step}, we may find that our influence function on the second term of the right side of the equation is also based on its empirical (sample) mean, rather than the true (population) expectation. The process of converging the empirical influence function towards the "true" influence function is called \textbf{empirical process}, as we decompose the difference between the one-step estimator and the true estimand, excluding the \(R_2\) terms which under our assumption is negligible: 
\begin{align*}
    \hat{\psi}(\mathbb{P}_n)_{\text{1-step}} - \psi(\tilde{\mathcal{P}}) &= \sum_{i=1}^n\phi(\psi; \mathbb{P}_n; Z_i) \\&= \sum_{i=1}^n\bigg[\left(\phi(\psi; \mathbb{P}_n; Z_i) -\phi(\psi; \tilde{\mathcal{P}}; Z_i)\right) + \left(\phi(\psi; \tilde{\mathcal{P}}; Z_i) - E_{\tilde{\mathcal{P}}}[(\phi(\psi; \tilde{\mathcal{P}}; Z) ]\right) \\& \quad+ \left(E_{tilde{\mathcal{P}}}[(\phi(\psi; \tilde{\mathcal{P}}; Z) ] - E_{\tilde{\mathcal{P}}}[(\phi(\psi; \mathbb{P}_n; Z) ]\right)\bigg] \\&=  \sum_{i =1}^n\phi(\psi; \mathbb{P}_n; Z_i) - E_{\tilde{\mathcal{P}}}\phi(\psi; \mathbb{P}_n; Z) ]
\end{align*}
As the second term represents the influence function from the true estimand, it should be zero due to the property of the influence function. This equation thus indicates that as the empirical measure \(\mathbb{P}_n\) converges to the true distribution \(\tilde{\mathcal{P}}\) the difference between the empirical influence function \(\phi(\psi; \mathbb{P}_n; Z_i)\) and the true influence function \(\phi(\psi; \tilde{\mathcal{P}}; Z)\) approaches zero\footnote{Recall that in our definition of the influence function for the empirical measure \(\mathbb{P}_n = \tilde{\mathcal{P}}_\epsilon\) approaches the influence function of the measure \(\tilde{\mathcal{P}}\) (which \(\epsilon = 1\)). This is consistent with our Figure \ref{fig:taylor} illustration on the Distributional Taylor Expansion, as we only have the derivative at \(\epsilon = 1\) end.}. In other words, the empirical influence function \(\phi(\psi; \mathbb{P}_n; Z_i)\), which serves as an estimator for the true influence function \(\phi(\psi; \tilde{\mathcal{P}}; Z)\), must also have the property of consistency \footnote{This is specifically \(L_2\) consistency.} as the sample size increases. This consistency ensures that the empirical influence function accurately reflects the true influence function as the number of samples grows, thereby contributing to the overall convergence of the one-step estimator to the true parameter.\\

With the influence function of the empirical data converging to the influence function of the estimated, we need statistical methods to control the speed of the convergence at \(1/\sqrt{n}\) so that our one-step estimator can be asymptotically valid and become the unbiased estimator of the true estimand in large sample size case. To ensure the convergence speed is no slower than \(1/\sqrt{n}\), we either need \textbf{sample splitting} techniques to process the original data, or we could assume that the empirical influence function falls into the \textbf{Donsker class}\footnote{A class of functions \(\Phi\) is a Donsker class if the empirical process indexed by \(\Phi\) converges in distribution to a Gaussian process: \(\mathbb{G}_n(\Phi) = \sqrt{n}(\mathbb{P}_n-\mathcal{P})\Phi\), as \(\mathbb{P}_n\) denotes the empirical measure and \(\mathcal{P}\) is the true underlying probability measure. The Donsker class has the property \(\mathbb{G}_n \rightsquigarrow \mathbb{G}\), converging to a Gaussian Process. In some ways, we could rewrite the empirical process as \(\sqrt{n}(\mathbb{P}_n-\mathcal{P})(\hat{\Phi}-\Phi)\), if we regard the convergence of the empirical measure \(\mathbb{P}_n\) and the convergence of the influence function \(\hat{\Phi}\) together at the rate of \(1/\sqrt{n}\). Obviously, the Donsker class property ensures that the empirical process converges uniformly and at a controlled rate. When the influence function of an estimator belongs to a Donsker class, this guarantees that the empirical process does not exhibit erratic behavior and converges smoothly.}. In this thesis, we only use sample-splitting techniques to ensure that our estimator is valid on the estimand.\\

Sample splitting is the most common technique to control the speed of the empirical process and make the efficient estimator a valid proxy for the true estimand. In machine learning, the method is also called \textbf{cross-validation}. Moreover, since we need to split and fit the model in at least two sub-datasets, this process is also called \textbf{double machine learning}, and the estimator with this strategy is the \textbf{double machine learning (DML)} estimator (intrinsically, it is the efficient estimator). \\
    
We can prove the effect of sample splitting on controlling the convergence rate of the empirical influence function towards the influence function from the estimand with Theorem \ref{mltclt}. To make the case simple, consider splitting the dataset \((Z_1,..., Z_n)\) only into two subsets: \(S_1 = \{Z_1, Z_2, \ldots, Z_{n_1}\}\) be the first subset, and \(S_2 = \{Z_{n_1+1}, Z_{n_1+2}, \ldots, Z_n\}\) be the second subset. \\

We first use \(S_1\) to construct an initial estimator \(\hat{\psi}_0\) of the parameter \(\psi\). Which, presumably, is the consistent estimator on the estimand \(\hat{\psi}\). We may write the one-step estimator with sample splitting as: 
\[
\hat{\psi}_{\text{1-step}} = \hat{\psi}_0 + \frac{1}{n_2} \sum_{i \in S_2} \phi(\hat{\psi}_0, \mathbb{P}_{n_2}, Z_i)
\]
Notice that for the naive estimator, we use the result from \(S_1\), while for the first-order bias correction term, we use estimators from \(S_2\) (as \(n_2\) is the size of the split \(S_2\), and \(\mathbb{P}_{n_2} = \frac{1}{n_2} \sum_{i \in S_2} \delta_{Z_i}\) is the empirical measure based on \(S_2\)). Obviously, the one-step estimator is consistent, as the law of large numbers indicates \(\hat{\psi}_0 \xrightarrow{prob.} \psi\) and the correction term\(\frac{1}{n_2} \sum_{i \in S_2} \phi(\hat{\psi}_0, \mathbb{P}_{n_2}, Z_i)\) will converge in probability to zero if \(\phi(\psi, \mathbb{P}_n, Z_i)\) is well-defined. Meanwhile, since the naive estimator and the first-order bias correction term from independent datasets, the construction of \(\hat{\psi}_0\) is irrelevant to the data used to make the correction.\\

Now we turn to the asymptotical normality. Due to the CLT, we have:
\[
\frac{1}{\sqrt{n_2}} \sum_{i \in S_2} \phi(\psi, \mathbb{P}_{n_2}, Z_i) \xrightarrow{d} \mathcal{N}(0, \sigma^2)
\]

Since \(\hat{\psi}_0 \xrightarrow{prob.} \psi\) and \(
\phi(\hat{\psi}_0, \mathbb{P}_{n_2}, Z_i) \approx \phi(\psi, \mathbb{P}_{n_2}, Z_i),
\)
we have: 
\[
\frac{1}{\sqrt{n_2}} \sum_{i \in S_2} \phi(\hat{\psi}_0, \mathbb{P}_{n_2}, Z_i) \xrightarrow{d} \mathcal{N}(0, \sigma^2) \quad \text{ (Slutsky's Theorem)}
\]
Therefore, the one-step estimator with sample splitting method is asymptotically normal with mean \(\psi\) and variance \(\sigma^2/n_2\), and the control on the convergence rate at \(1/\sqrt{n}\) is accomplished. \\

In summary, with the second (and higher) order remainders and the empirical process becomes insignificant under the large sample size scenario, we can use the one-step estimator: the empirical naive estimator plus the first-order bias correction term to robustly and efficiently infer the true estimand. 

\subsection{Efficient Influence Functions and Efficient Estimators}
\subsubsection{Efficient influence functions}
So far, we have discussed the score function, the influence function, and the estimator from the general RAL conditions. With the tools we have, we could have proof of the CRB brought up at the start of this section: \\

\begin{proof}[Proof on Cramer-Rao Bound]
    Since our estimator \(\hat{\psi} = \psi(\mathbb{P}_n)\) is an unbiased estimator of \(\psi = \psi(\mathcal{P})\), therefore, \(E[\hat{\psi}] = \psi\). Our target is to get the upper bound for the variance of \(\hat{\psi}\). Recall that we could use Cauchy inequality to grab the upper bound for the variance if the expectation is known: \(E[XY]^2 \leq E[X]^2E[Y]^2\), we use the score function pointing to \(\psi\) (from \(\hat{\psi}\)) as the auxiliary function, as the expectation of the score function is zero. We denote the score function as \(s(Z; \psi) = \nabla_{\psi}(\tilde{p}(z) = \frac{\partial \log \tilde{p}_\epsilon(z)}{\partial \epsilon }\big|_{\epsilon = 0}\) to show the effect of \(\psi\). Therefore, 
    \[
    cov(\hat{\psi}; S(Z; \psi)]^2 \leq \text{Var}(\hat{\psi})E[S(Z; \psi)]^2 \Rightarrow \text{Var}({\hat{\psi}}) \geq \frac{cov(\hat{\psi}; s(Z;\hat{\psi}))}{E[S(Z; \psi)^2]}
    \]
    Since we define the score function on the direction to \(\hat{\psi}\), the two components in the covariance \(\hat{\psi}\) and \(s(Z;\psi)\) are indeed in the same direction, and therefore the covariance is \(1\), we denote the expectation of the square of the score function as the Fisher information: \(I(\psi) = E[s(Z; \psi)^2]\); therefore, we have the Cramer-Rao bound for the variance of the unbiased estimator \(\hat{\psi}\):
    \[
    \text{Var}({\hat{\psi}}) \geq \frac{1}{I(\psi)}
    \]
\end{proof}

As we noted earlier in this chapter, if the estimator \(\hat{\psi}\) attains the Cramer-Rao bound, we call the estimator the efficient estimator. Meanwhile, its associated influence function is called the \textbf{effient influence function (EIF)}. We can either have the EIF to get the efficient estimator, or reversely obtain the efficient estimator and get its EIF. Usually, we adopt the first strategy: we first find a set \(\Phi\) of possible influence functions and select the most efficient one which satisfies \(\phi^\dagger = \arg\min_{\phi \in \Phi}\text{Var}[\phi]\)\footnote{It's not hard to understand that the lowest variance for the estimator indicates the lowest variance (square mean) for the influence function so I didn't include it in the main text: we could just get it from Equation \ref{AL3} on the definition of asymptotical linearity; or we could still use the Taylor Distributional expansion to understand it, as \(\hat{\psi}\) is an unbiased estimator for \(\psi\), so the variance only remain in the higher-order derivatives and the increase in variance leads to the increase in the derivatives, which leads the increase in the variance in the influence functions.} and we finally get the efficient estimator. \\

A crucial property for the EIF \(\phi^\dagger\) is that it lies in the tangent space. This is because only the influence function in the tangent space obtains the lowest variance: \(\phi^\dagger \perp h^{\perp}\), where \(h^{\perp}\) stands for the lines orthogonal to any element in the tangent space \(h \in \mathcal{T}\). \\

\begin{proof}
    We suppose we could decompose the influence function into two orthogonal parts: one is the projection on the tangent space \(h\), and one is orthogonal to the tangent space, \(\phi-h\). For any score function \(s\) on the tangent space, according to the definition, we have: \(E[(\phi-h)s] = 0\) \footnote{The influence function and the score function are both Hilbert space (since they are both \(L_2^0\) space as we mentioned before, and we could also use the inner product of Hilbert space to represent it, \(\langle \phi-h, s \rangle = 0, \forall s \in \mathcal{T}(\mathcal{P})\). The orthogonal decomposition is justified by the Reisz Representation Theorem, which states that for every continuous linear functional \(\mathcal{F}\) on the Hilbert space \(\mathcal{H}\), there is always an element \(h in \mathcal{H}\) such that \(\mathcal{F}(f,h) = \langle f, h \rangle\) \(\forall f \in \mathcal{H}\). }. Thus, for the variance of the influence function: 
    \begin{align*}
        \mathrm{Var}(\phi) &= \mathrm{Var}(h + (\phi-h)) = \mathrm{Var}(h) + \mathrm{Var}(\phi-h) \text{ (orthogonality, no covariance)} \\ &= E[h^2]-(E[h])^2 + E[(\phi-h)^2] + (E[\phi-h)])^2 \\ &= E[h^2] + E[(\phi-h)^2] 
    \end{align*}
    Since \(h\) is constant (as it is the projection of the influence functions onto the tangent space), the target function \(\phi^\dagger = \arg \min_{\phi}(\phi-h)^2 \Rightarrow \phi^\dagger = h\). Therefore, the influence function should lie on the tangent space to be efficient.
\end{proof}
Therefore, we call the EIF the "\textbf{canonical gradient}." With this property or the EIF, we have another direct corollary: for the saturated models, there is one influence function, which is the EIF. This is simply because, the score functions of the saturated model point to all directions and therefore could take any values satisfying zero mean and finite variance. Suppose we have two influence functions \(\phi_1\) and \(\phi_2\), therefore, \(E[\phi_1 s] = E[\phi_2 s] = E[(\phi_1 - \phi_2)s = 0] \Rightarrow \phi_1 = \phi_2\). However, this is not the case for the non-saturated models, as the restrictions of the non-saturated models make some directions orthogonal to the tangent space possible. \\

\subsubsection{Derving EIF for the ATE}
The review above in this section gives a rough (but lengthy) introduction to the efficient theory and the efficient estimator, and all the estimators in this thesis are efficient (whose estimation errors reach the CRB or, at least, the most efficient one). The remaining job in the introductory chapter is to derive the efficient estimator for the average treatment effect, which we have discussed in Section \ref{sec2}: \(Y(1) - Y(0)\). \\

We start with the saturated model: the ATE in the observational study, where there is no restriction on the statistical estimand to infer the causal estimand, and there's only one influence function-- which is the EIF for the estimator. Similar to the operation in Equation \ref{expectation}, we use \(\psi_a = E[E_X[Y | A= a, X]]\) from the observational study to infer \(\psi_a^* = E[Y(a)]\) (therefore, the ATE is \(\psi_1 - \psi_0)\). From the perspective of the efficient theory, the estimators we applied in Section \ref{sec2}, for instance, the IPW estimator, as we have shown, is a RAL estimator but not the efficient one, as it is only the "naive plug-in estimator" part in the Taylor distributional expansion decomposition (Equation \ref{eq:taylordecomp}). Our goal is the efficient estimator for \(\psi_a \):
\[
\psi_a = E[E_X[Y | A= a, X]] = \sum_x \underbrace{E[Y| A = a, X]}_{:=\mu_a(X)}p(x) = \sum_x \mu_a(x)p(x)
\] 
As we discussed above, to derive the efficient estimator, we first capture the EIF and then derive the efficient estimator based on the EIF. We need to calculate 
\begin{equation} \label{eq:eif1}
\phi(\psi_a) = \sum \phi(\mu_a(x)p(x)) = \sum\bigg(\phi(\mu_a(x))p(x) + \mu_a(x)\phi(p(x))\bigg).     
\end{equation}

As the equation shows, we need the EIFs for the expectation \(\phi(E[X])\) (to yield \(\phi(p(x))\))and the conditional expectation \(\phi(E[Y|X = x])\) (to yield \(\phi(\mu_a(x)\)).  \\

We first derive the EIF for \(\psi(\mathcal{P}) = E_\mathcal{P}[X] = \int x d\mathcal{P}(x).\) We start with the Gateaux derivative definition of the influence function. The perturbed distribution is set as: 
\[
\tilde{\mathcal{P}}_\epsilon = (1- \epsilon)\mathcal{P} + \epsilon\delta_x
\] 
Which is similar to what we did in Equation \ref{derivative}. Therefore, we construct the nominator \(\psi(\tilde{\mathcal{P}}_\epsilon)\):
\begin{align*}
\psi(\tilde{\mathcal{P}}_\epsilon) &= \int x 
 d\tilde{\mathcal{P}}_\epsilon = \int x d\big((1-\epsilon)\mathcal{P} +\epsilon \delta_x\big) \\&= (1-\epsilon)\int x d\mathcal{P} + \epsilon \int x d \delta_x \\&= (1-\epsilon)\psi(\mathcal{P}) + \epsilon x \\& = \psi(\mathcal{P}) + \epsilon(x - \psi(\mathcal{P}))
\end{align*}
Since \(\int d\delta_x = 1\). Thus, we have the influence function \(\phi(\psi = E_{\mathcal{P}}[X];\mathcal{P}, x)\) :
\begin{equation}\label{eq:ifexp}
\phi(E_{\mathcal{P}}[X]) = \frac{\partial}{\partial \epsilon}\psi(\tilde{\mathcal{P}}_\epsilon)\big|_{\epsilon = 0} = x - \psi(\mathcal{P}) = x - E_{\mathcal{P}}[X]    
\end{equation}

We then derive the EIF for \(\psi(\mathcal{P}) = E_{\mathcal{P}}[Y |X = x] = \int_y y d\mathcal{P}(y |x)\). Still, \(\tilde{\mathcal{P}}_\epsilon = (1- \epsilon)\mathcal{P} + \epsilon\delta_{y|x}\). The hard part we need to derive here is \(\mathcal{P}(y |x)\). We may recall the Bayesian rule \(\mathcal{P}(y |x) = \frac{\mathcal{P}(y,x)}{\mathcal{P}(x)}\). Therefore, 
\begin{align*}
  \tilde{\mathcal{P}}_\epsilon(y |x) = \frac{\tilde{\mathcal{P}}_{\epsilon}(y,x)}{\tilde{\mathcal{P}}_{\epsilon}(x)} = \frac{(1-\epsilon)\mathcal{P}(y,x) + \epsilon \delta_{y,x}}{(1-\epsilon) \mathcal{P}(x) + \epsilon\delta_x}  
\end{align*}
Given by the equation above. Since the influence function is defined as \(\frac{\partial\psi(\tilde{\mathcal{P}}_\epsilon)}{\partial\epsilon}\big|_{\epsilon = 0} \) is a gradient, we first deal with the part to be integrated (\( d\tilde{\mathcal{P}}_{\epsilon}(y|x)\)), which we have transformed with the above equation. Recall the gradient algebra rule: for two functions \(u\) and \(v\), \(\big(\frac{u}{v} \big)'= \frac{u'v-uv'}{v^2}\). Therefore, 
\begin{align*}
   \tilde{\mathcal{P}}_{\epsilon}(y|x)\big|_{\epsilon = 0} &= \frac{[(1-\epsilon)\mathcal{P}(y,x) + \epsilon \delta_{y,x}]'[(1-\epsilon) \mathcal{P}(x) + \epsilon\delta_x] - [(1-\epsilon)\mathcal{P}(y,x) + \epsilon \delta_{y,x}][(1-\epsilon) \mathcal{P}(x) + \epsilon\delta_x]'}{[(1-\epsilon) \mathcal{P}(x) + \epsilon\delta_x]^2} \\&= \frac{[\delta_{y,x}-\mathcal{P}(y,x)][(1-\epsilon)\mathcal{P}(x)+\delta_x]-[\delta_x-\mathcal{P}(x)][(1-\epsilon)\mathcal{P}(y,x) + \delta_{y,x}]}{[(1-\epsilon) \mathcal{P}(x) + \epsilon\delta_x]^2} \\&=\frac{\mathcal{P}(y,x)\delta_x - \delta_{y,x}\mathcal{P}(x)}{\mathcal{P}(x)^2} \text{ (since }\epsilon = 0)
\end{align*}
Therefore, The EIF is: 
\begin{align*}
    \frac{\partial\tilde{\mathcal{P}}_\epsilon}{\partial\epsilon}\big|_{\epsilon = 0} &= \int y d\bigg[\frac{\mathcal{P}(y,x)\delta_x - \delta_{y,x}\mathcal{P}(x)}{\mathcal{P}(x)^2} \bigg] \\
    & = \frac{\delta_x}{p(x)}\bigg[\int y d\frac{\delta_{y,x}}{\delta_x} - \int y d\frac{\mathcal{P}(y,x)}{\mathcal{P}(x)}\bigg] \\&= \frac{\mathbbm{1}_{x}}{p(x)}\bigg[\int y d \delta_{y|x} - \int y d \mathcal{P}(y| x)\bigg] \text{ (Since }\frac{\mathcal{P}(y,x)}{\mathcal{P}(x)} = \mathcal{P}(y|x); \int d \delta = \mathbbm{1})\\ & = \frac{\mathbbm{1}_x}{p(x)}[y - E_{\mathcal{P}}[Y | X = x]]
\end{align*}
We get:
\begin{equation}\label{eq:ifcon}
    \phi(\psi = E_{\mathcal{P}}[Y |A = a, X = x], \mathcal{P}, (y,a, x)) = \frac{\mathbbm{1}_{(a, x)}}{p[A= a, x]}\bigg[y - E_{\mathcal{P}}[Y | A = a, X]\bigg]
\end{equation}
Which implies that \(\phi(\mu_a(x)) \). With Equations \ref{eq:eif1}, \ref{eq:ifexp} and \ref{eq:ifcon}, we have the EIF for the counterfactual estimand \(\phi(\psi_a) = E[E_X[Y | A = a, X ]],\mathcal{P}, (y,a,x))\). Since \(\mu_a(x) = E[Y | A = a, X = x]\), \(\phi(\mu_a(x)) = \frac{\mathbbm{1}(a,x)}{p(a,x)}\big[y - \mu_a(x)\big]\); \(p(x) = p(x)\), \(\phi(p(X)) = \mathbbm{1}_x - p(x)\); and \(\pi_a(X) = p(A = a|X)\), \(\psi_a = E[Y |A= a]\). Therefore, we have:
\begin{align*}
    \phi(\psi_a) &= \sum_x \bigg[\big(\frac{\mathbbm{1}_{a, x}}{p(a,x)}[y - \mu_a(x)]p(x)\big) + \big(\mu_a(x)(\mathbbm{1}_x - p(x))\big)\bigg]  \\&= \frac{\mathbbm{1}(a)}{\pi_a(x)}[y - \mu_a(x)] + \mu_a(x) - \psi_a \\& =
    \frac{\mathbbm{1}(A = a)}{p(A = a | X)} (Y - E[Y | A = a, X]) + E[Y | A = a, X] - E[Y | A = a] 
\end{align*}
Thus, the EIF for \(\psi_a\) with discrete measurable elemtes \(\phi(\psi_a, \mathcal{P}, (Y_i, X_i, A_i))\) is: 
\begin{equation}\label{eq:eifpart}
    \phi(\psi = \psi(\mathcal{P}), \mathcal{P}, (Y_i, X_i, A_i)) = \frac{A_i}{\pi_a(X_i)} [(Y_i - \mu_{a}(X_i)] + \mu_{a}(X_i) - \psi_{a}
\end{equation}
In other words, 
\[
    \phi(\psi_1) =  \frac{\mathbbm{1}(A_i = 1)}{\pi_1(X_i)} (Y_i - \mu_1(X_i)) + \mu_1(X_i) - \psi_1
\]
and \[
\phi(\psi_0) =  \frac{\mathbbm{1}(A_i = 0)}{\pi_0(X_i)} (Y_i - \mu_0(X_i)) + \mu_0(X_i) - \psi_0
\]
We could use the central identity for the influence function (Equation \ref{central}) to verify if our EIF is correct. \\

\begin{proof}
 Suppose we are verifying the EIF for \(\psi_0\). Using the factorizing of the tangent space technique, we could decompose the gradient of the estimator in any direction into:
\[
\nabla_s \psi_0 = \nabla_{s_{Y|A,X}}\psi_0 + \nabla_{s_{A|X}}\psi_0 + \nabla_{s_{X}}\psi_0
\]
Similarly, for the expectation of the influence and the score function, we could also decompose it through the algebra of expectations:
\[
E[\phi s]= E[\phi s_{Y|A,X}] + E[\phi s_{A|X}] + E[\phi s_{X}]
\]
Therefore, we need to prove that \(\nabla_{s_{Y|A,X}}\psi_0 = E[\phi s_{Y|A,X}] \), \(\nabla_{s_{A|X}}\psi_0 = E[\phi s_{A|X}] \), and \( \nabla_{s_{X}}\psi_0 = E[\phi s_{X}]\), correspondingly. 
We begin with \( \nabla_{s_{X}}\psi_0 = E[\phi s_{X}]\). 
\begin{align*}
 \nabla_{s_{X}}\psi_0 &= \frac{\partial}{\partial\epsilon} \int_x \int_y \tilde{p}_\epsilon(y| 0, x) dy \tilde{p}_\epsilon(x) dx =   \int_x\int_y p(y |0,x) dy \frac{\partial}{\partial \epsilon}(1+ \epsilon s) p(x) dx \\&= \int_x\int_y p(y |0,x) dy s_X p(x) dx = \int_x \mu_0(x)s_X p(x)dx   
\end{align*}
And,
\begin{align*}
  E[\phi s_{X}] & = E\left[\left(\frac{\mathbbm{1}(A = 0)}{\pi_0(x)}[y - \mu_0(x)] + \mu_0(x)\right) s_X \right] \\&= \int_x \int_y \left[ \left(\frac{\mathbbm{1}(A= 0)}{\pi_0(x)}[y - \mu_0(x)] + \mu_0(x)\right) s_X \right] p(y|x,a) \, dy \, p(x) \, dx \\&=\int_x s_X\left[\int_y \frac{\mathbbm{1}(A = 0)}{\pi_0(x)}[y - \mu_0(x)] p(y|x,a) \, dy + \mu_0(x) \int_y p(y|x,a) \, dy  \right] p(x) \, dx 
\end{align*}
Notice that \(\int_y p(y|x,a)dy = 1\) and \(\int_y[y-\mu_0(x)]p(y|x,a)dy = E[y-\mu_0(x)|x,0] = 0\), thus, 
\begin{align*}
    E[\phi s_{X}] = \int_x s_X\mu_0(x)p(x)dx = \nabla_{s_X} \psi_0
\end{align*}
Similarly, 
\begin{align*}
 \nabla_{s_{A|X}}\psi_0 &= \int_x\int_y p(y \mid 0,x) dy \frac{\partial}{\partial \epsilon}(1+\epsilon s_{A|X})p(x)d(x) \\&= \int_x\mu_0s_{A|X}p(x)dx = 0  
\end{align*}
\begin{align*}
  E[\phi s_{A|X}] &= \int_x \int_a \phi s_{A|X}
(a|x)p(a|x)p(x)dadx \\&= \int_x\phi \underbrace{\bigg(\int_a s_{A|X}(a|x)p(a|x)da\bigg)}_{ = 0}p(x)dx \\&= 0 = \nabla_{s_{A|X}}\psi_0 
\end{align*}
At last, 
\begin{align*}
    \nabla_{s_{Y|A,X}} \psi_0 &= \int_x\int_yp(y|x, 0) s_{Y |A,X}dy p(x)dx \\& = \int_xp(x)\bigg(\int_y p_0(y |x)s_{Y|A,X}dy\bigg)dx \\& = \int_x (y|A =  0)s_{Y|A,X}dx
\end{align*}
\begin{align*}
    E[\phi s_{Y|A,X}] &= \int_x \int_y \left[ \left(\frac{\mathbbm{1}(A= 0)}{\pi_0(x)}[y - \mu_0(x)] + \mu_0(x)\right) s_{Y|A,X} \right] p(y|x,a) \, dy \, p(x) \, dx \\&
    = \int_x \bigg[\frac{\mathbbm{1}(A= 0)}{\pi_0(x)}y s_{Y|A,X}\bigg]p(x)dx \\& = \int_x (y| A= 0)s_{Y|A,X}p(x)dx = \nabla_{s_{Y|A,X}}\psi_0.
\end{align*}
Therefore, \(E[\phi s] = \nabla_s \psi\) and our derivation on the EIF is correct. 
\end{proof}

Further, we consider the non-saturated model scenario. The average treatment effect under the random-controlled trial setting is a non-saturated model since we have placed restrictions on the treatment and control cases. Therefore, unlike the saturated models, we cannot derive the efficient influence function with an arbitrary score function, as the influence function may not be efficient. However, we could start with a known RAL estimator and derive its influence function through the definition of asymptotical linearity (Equation \ref{AL} instead of the Gateaux derivative definition 
 in Equation \ref{derivative}), and then project it (find its minimized square error) onto the tangent space. As elaborated before, the EIF should be the projection of any influence functions on the tangent space. However, it might be hard to directly find the projection on the tangent space if the underlying estimator is complex. If so, we could still use the factorization technique that first projects the influence function onto tangent subspaces and then sums the factorized influence functions together. \\

For the ATE under the RCT setting, we start with the IPW estimator. Obviously, as shown in Equation \ref{weights}, the IPW estimator is a regular and asymptotical linear one. We still use the estimation on the EIF for \(\psi_0\) as an example below: 
\[
\hat{\psi}_0^{IPW} = E\left[\frac{\mathbbm{1}(A_i=0)}{\pi_0(X_i)}Y_i\right] = \sum_{i=1}^n \left[\frac{\mathbbm{1}(A_i=0)}{\pi_0(X_i)}Y_i\right].
\]
Therefore, based on Equation \ref{AL}, we could derive the influence function as:
\[
\phi_0^{IPW} = \frac{\mathbbm{1}(A_i=0)}{\pi_0(X_i)}Y_i - \psi_0.
\]
Then we project the influence function from the IPW estimator into the tangent subspace of \(\mathcal{T}_{Y|A, X}\), \(\mathcal{T}_{A|X}\), and \(\mathcal{T}_{X}\) in which \(s_{Y|A, X}\), \(s_{A|X}\), and \(s_{X}\) forms (and \(\mathcal{T}_{Y, A, X} = \mathcal{T}_{Y|A, X} \oplus \ \mathcal{T}_{A|X} \oplus \mathcal{T}_{X}\). \\

\begin{figure}[ht]
    \centering
    \includegraphics[width=\textwidth]{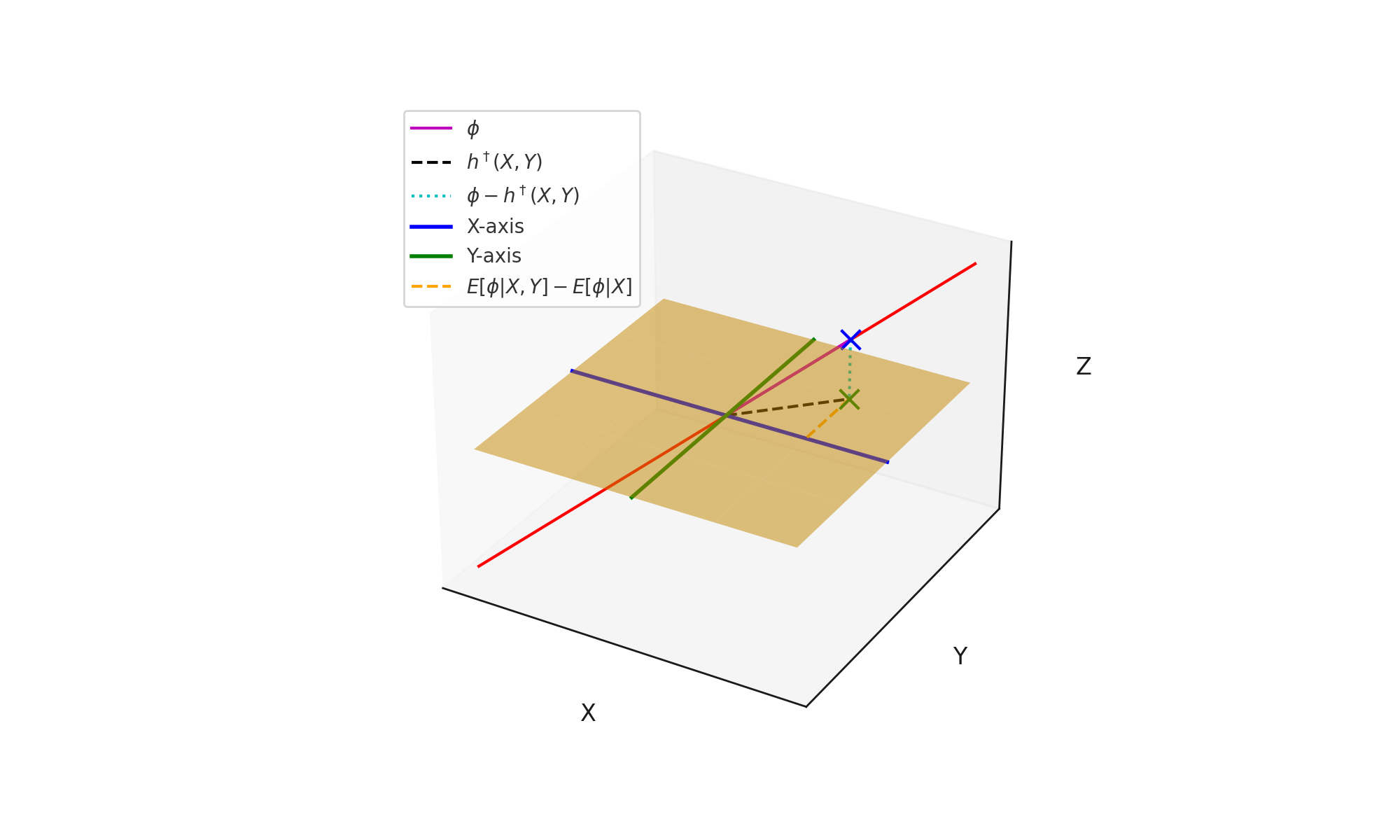}
    \caption{Illustrations on projections of influence functions}
    \label{fig:proj}
    
    \raggedright \small \textbf{Note:} The yellow plane denotes the tangent space. The figure shows projecting the (inefficient) influence function (not on the tangent space) towards the joint distribution of \((X, Y)\) and the marginal distribution of \((Y |X = 0)\).
\end{figure}

First, we try to project the influence function of the IPW estimator on the tangent space \(\mathcal{T}_{X}\). The projection function is defined to find the score function on the tangent space for which its mean square error with the influence function from the IPW is minimal\footnote{We have a sketch Figure \ref{fig:proj} illustrating the projection process for the readers' reference for understanding the algebraic process here.}. For any influence function, 
\[
Proj_{\mathcal{T}_X}(\phi) = \arg \min_{h(X) \in \mathcal{T}_X} E[(\phi - h(X)^2].
\]

Let \(h^\dagger(X) = \arg \min_{h(X) \in \mathcal{T}_X} E[(\phi - h(x)^2] \), therefore, the residual \(\phi - h^\dagger(X)\) should be orthogonal to functions on \(\mathcal{T}_X\):
\[E[(\phi - h^\dagger(X))h(X)] = 0, \, \forall h(X) \in \mathcal{T}_X.\]
Since \(h^\dagger(X)\) satisfies the orthogonality condition, it should be the conditional expectation of \(\phi\) given \(X\):
\[
h^\dagger(X) = E[\phi | X].
\]
Therefore, we have the projection of the influence function for the IPW estimator on the tangent space \(\mathcal{T}_X\) as its conditional expectation on the \(X\) axis: 
\begin{align*}
 \phi^{\dagger}_{0 \langle \mathcal{T}_X \rangle} &= E[\phi_0^{IPW} \mid X] = E\left[\left(\frac{\mathbbm{1}(A_i = 0)}{\pi_0(X_i)}Y_i - \psi_0\right) \mid X \right] \\&= E\left[E\left[\frac{\mathbbm{1}(A_i =0)}{\pi_0(X_i)}Y_i \mid A_i, X\right ] \mid X\right] -\psi_0 \\& = E\left[E\left[\frac{Y_i}{\pi_0(X_i)}\mid A_i = 0, X\right]\mathbbm{1}(A_i = 0) \mid X\right] - \psi_0 \\& = E\left[Y_i | A_i = 0, X\right] - \psi_0 = \mu_0(X_i) - \psi_0 
\end{align*}

Now we derive the projection of the influence function to the conditional tangent space. If we project the influence function onto the plane spanned by two vectors \(X\) and \(Y\), \(\mathcal{T}_{X, Y}\), obviously, the projection should be the conditional expectation of the score function conditioned on the joint distribution of \(X, Y\):
\[
h^\dagger(X, Y) = E[\phi | X, Y]
\]
And if we project an influence function towards the plane spanned by \(X\) and \(Y\) but conditioned on the line which \(X = 0\), then the projection should be the above result, minus its projection on the \(X\)-axis, which is:
\[
h^\dagger(Y |X = 0) = E[\phi | X, Y] - E[\phi |X]
\]
We denote the tangent space in this situation \(\mathcal{T}_{Y, X_0}\). Now we discuss the projections of the influence function of the IPW estimator onto the tangent spaces \(\mathcal{T}_{A_0 |X}\) and \(\mathcal{T}_{Y|A_0, X}\). Obviously, 
\[
\phi^{\dagger}_{0 \langle \mathcal{T}_{A_0 |X} \rangle} = 0,
\]
as the tangent space \(\mathcal{T}_{A_0 |X}\)is orthogonal to \(\phi_0^{IPW}\) and therefore all projections have no length. For the projection of \(\phi_0^{IPW}\) onto the tangent space \(\mathcal{T}_{Y |A_0, X}\), we have:
\begin{align*}
    \phi^\dagger_{0\langle\mathcal{T}_{Y|A_0, X}\rangle} &= E[\phi_0^{IPW} | Y, A_0, X] - E[\phi_0^{IPW} | A_0, X] \\& =E\left[\frac{\mathbbm{1}(A_i) = 0}{\pi_0(X_i)}Y_i - \psi_0 \mid Y, A_0, X \right] - E\left[\frac{\mathbbm{1}(A_i = 0)}{\pi_0(X_i)}Y_i - \psi_0 \mid \ A_0, X \right] \\& = \left(\frac{\mathbbm{1}(A_i = 0)}{\pi_0(X_i)}Y_i - \psi_0 \right) - \left(\frac{\mathbbm{1}(A_i) = 0}{\pi_0(X_i)}\mu_0(X_i) - \psi_0  \right)
\end{align*}
Therefore, we sum up the three sub-EIFs on the three tangent subspaces and get the EIF for \(\psi_0\) under the RCT settings:
\begin{align*}
    \phi^\dagger(\hat{\psi}_0^{IPW}) \\&=  \phi^{\dagger}_{0 \langle \mathcal{T}_X \rangle} +\phi^{\dagger}_{0 \langle \mathcal{T}_{A_0 |X} \rangle}+\phi^\dagger_{0\langle\mathcal{T}_{Y|A_0, X}\rangle} \\&=  \left(\mu_0(X_i) - \psi_0\right) + 0 + \left(\frac{\mathbbm{1}(A_i = 0)}{\pi_0(X_i)}Y_i - \psi_0 \right) - \left(\frac{\mathbbm{1}(A_i) = 0}{\pi_0(X_i)}\mu_0(X_i) - \psi_0  \right) \\& = \frac{\mathbbm{1}(A_i = 0)}{\pi_0(X_i)} (Y_i - \mu_0(X_i)) + \mu_0(X_i) - \psi_0
\end{align*}
Which is, unsurprisingly, exactly the EIF for \(\psi_0\) we obtained from the saturated model. Similarly, the EIF for \(psi_1\) from the non-saturated model of the RCT will also be the same as the EIF from the saturated model of the observational study. \\

With the EIFs for \(\psi_1\) and \(\psi_0\), we can get the EIF for the ATE:
\begin{align}
    \phi(\psi) &= \phi(\psi_1) - \phi(\psi_0) \notag\\
    & = \bigg[\frac{\mathbbm{1}(A_i = 1)}{\pi_1(X_i)} (Y_i - \mu_1(X_i)) + \mu_1(X_i) - \psi_1 \bigg] - \bigg[ \frac{\mathbbm{1}(A_i = 0)}{\pi_0(X_i)} (Y_i - \mu_0(X_i)) + \mu_0(X_i) - \psi_0\bigg] \notag \\& = \frac{\mathbbm{1}(A_i = 1)}{\pi(X_i)}\left(Y_i - \mu_1(X_i)\right) - \frac{1-\mathbbm{1}(A_i = 1)}{1-\pi(X_i)}\left(Y_i - \mu_0(X_i)\right) + \left(\mu_1(X_i) - \mu_0(X_i) \right) - \psi
\end{align}
As we let \(\pi(X_i) = \pi_1(X_i) = \mathcal{P}(A_i = 1 |X_i)\) and \(\psi_1 - \psi_0 = \psi\). With Equation \ref{AL} at the start of this section, we could derive the efficient estimator for the average treatment effect:
\begin{equation}\label{eq:drestimator}
    \hat{\psi} = \frac{1}{n}\sum_{i= 1}^n \bigg[\frac{\mathbbm{1}(A_i = 1)}{\hat{\pi}(X_i)}\left(Y_i - \hat{\mu_1}(X_i)\right) - \frac{1-\mathbbm{1}(A_i = 1)}{1-\hat{\pi}(X_i)}\left(Y_i - \hat{\mu_0}(X_i)\right) + \left(\hat{\mu}_1(X_i) - \hat{\mu}_0(X_i) \right)\bigg]
\end{equation}
Equation \ref{eq:drestimator} is the core equation in the whole thesis. We call the estimator in Equation \ref{eq:drestimator}
in several ways. It is, as we elaborated, the efficient causal estimator, and it is also the \textbf{doubly robust (DR)} causal estimator since the estimator will be consistent as either our specification of \(\pi(X_i)\) or our specification of \(\mu_a(X_i)\) is correct. Further, it is also called the \textbf{debiased machine learning (DML) estimator} as it eliminates the bias between the estimator and the true estimand with the double robustness we will give the details below. Meanwhile, as we mentioned earlier, since cross-validation will be used in the algorithm generating the estimator to control the convergence speed of the empirical process, the estimator is also called \textbf{double machine learning (DML)} estimator. Finally, some literature also calls this estimator the \textbf{Neyman-orthogonal estimator} of the ATE, as the EIF satisfies the Neyman orthogonality, yielding this result. \\

From the perspective of debiased estimation, when the propensity score function \(\pi(X_i)\) is correctly specified, we have:
\[
\mathbb{E}\left[\frac{A_i}{\pi(X_i)} (Y_i - \mu_1(X_i)) \mid X_i \right] = 0
\]
and
\[
\mathbb{E}\left[\frac{1 - A_i}{1 - \pi(X_i)} (Y_i - \mu_0(X_i)) \mid X_i \right] = 0.
\]

Therefore, the first two terms of the influence function become mean-zero conditional on \(X\), leaving:
\[
\hat{\psi}_{\text{DR}} = \frac{1}{n} \sum_{i=1}^n \left[\mu_1(X_i) - \mu_0(X_i) \right].
\]
So even if \(\mu_1(X_i)\) and \(\mu_0(X_i)\) are misspecified, the terms involving the propensity score correct the bias introduced by the misspecified outcome models, resulting in a consistent estimator for the ATE. \\

Similarly, if the outcome regression models \(\mu_1(X)\) and \(\mu_0(X)\) are correctly specified, we have:
\[
\mathbb{E}[Y_i \mid A_i = 1, X_i] = \mu_1(X_i) \quad \text{and} \quad \mathbb{E}[Y_i \mid A_i = 0, X_i] = \mu_0(X_i).
\]

In this case, the terms \(Y_i - \mu_1(X_i)\) and \(Y_i - \mu_0(X_i)\) are mean-zero conditional on \(A_i\) and \(X_i\). Thus, the first two terms of the influence function average out to zero:
\[
\hat{\psi}_{\text{DR}} = \frac{1}{n} \sum_{i=1}^n \left[\frac{A_i}{\pi(X_i)} (Y_i - \mu_1(X_i)) - \frac{1 - A_i}{1 - \pi(X_i)} (Y_i - \mu_0(X_i)) + \mu_1(X_i) - \mu_0(X_i) \right].
\]

Here, even if \(\pi(X_i)\) is misspecified, the correctly specified outcome regression models ensure that the estimator is consistent for the ATE.\\

From the perspective of the convergence rate, we will find that the estimator also satisfies double robustness with its higher-order Distributional Taylor Expansion. According to Equation \ref{eq:taylordecomp}, We can derive the second-order remainder of the estimator \(\psi_a\) :
\begin{align*}
R_2 &= \psi(\tilde{\mathcal{P}}_\epsilon) - \psi(P) + \frac{\partial}{\partial \epsilon}\psi(\tilde{\mathcal{P}}_\epsilon)\big|_{\epsilon = 1} \\&= \psi(\tilde{\mathcal{P}}_\epsilon) - \psi(P) + \frac{1}{n}\sum_{i=1}^n \phi(\psi(\tilde{\mathcal{P}}_\epsilon)) \\&= \hat{\psi}_a - E[\mu_a(X_i)] + E\left[\frac{\mathbbm{1}(A_i = a)}{\hat{\pi}_a(X_i)}(Y_i-\hat{\mu}_a(X_i)) + \hat{\mu}_a(X_i)\right] -E\left[\hat{\psi}(a)\right] \\& = E\left[\frac{\pi_a(X_i)}{\hat{\pi}_a(X_i)}\mu_a(X_i) - \mu_a(X_i)\right] + E\left[\hat{\mu}_a(X_i) -\frac{\pi_a(X_i)}{\hat{\pi}_a(X_i)}\hat{\mu}_a(X_i)\right] \\&= E\left[\frac{1}{\hat{\pi}_a(X_i)}\left(\pi_a(X_i) -\hat{\pi}_a(X_i)\right)\left(\mu_a(X_i) - \hat{\mu}_a(X)\right)\right]
\\&\leq ||\pi_a(X_i) - \hat{\pi}_a(X_i)|| \,||\mu_a(X_i) - \hat{\mu}_a(X_i)|| \text{ (Cauchy-Schwarz Inequality)}
\end{align*}
We could see from this formation that either \(\pi_a(X_i) = \hat{\pi}_a(X_i)\) or \(\mu_a(X_i) - \hat{\mu}_a(X_i)\), the second-order remainder turns to zero. Therefore, either \(\hat{\pi}_a(X_i)\) or \(\hat{\mu}_a(X_i)\) is correctly specified, our estimator is unbiased. \\

Further, this inequality also suggests the convergence rate of the doubly robust estimator. Since we require the second-order remainder to converge at the rate of \(o_p(n^{-1/2})\), we could require both \(||\pi_a(X_i) - \hat{\pi}_a(X_i)||\) and \(||\mu_a(X_i) - \hat{\mu}_a(X_i)|| \) to converge at the rate of \(o_p(n^{-1/4})\). Regularized machine learning methods and cross-validation need to be used for both the propensity score model \(\hat{\pi}_a(X_i)\) and the outcome model \(\hat{\mu}_a(X_i)\). For instance, we could choose gradient boosting machines or random forests to predict the propensity and generalized additive models (GAM) or random forests to predict the outcome\footnote{A pretty useful programming package for model choice is called super-learner\citep{SuperLearner}, which could be convenient for social scientists choose the appropriate machine learning models for model fitting.}, and then use regularized techniques like lasso, ridge, or elastic nets to control the complexity of the models. \\

So, practically, we have the algorithm deriving the doubly robust (efficient) causal estimator:  
\begin{enumerate}

    \item \textbf{Set up Input:}
    \begin{itemize}
        \item Dataset \( \{(X_i, A_i, Y_i)\}_{i=1}^n \), where \(X_i\) represents covariates, \(A_i\) represents treatment assignment (0 or 1), and \(Y_i\) represents outcomes.
        \item Number of folds for cross-validation \(k\).
    \end{itemize}

    \item \textbf{Split Dataset for Cross-Validation}
    \begin{itemize}
        \item Randomly split the dataset into \(k\) approximately equal-sized folds. Each fold will be used as a validation set while the remaining \(k-1\) folds will be used for training.
        \item Label these folds as \( \{D_1, D_2, \ldots, D_k\} \).
    \end{itemize}

    \item \textbf{Cross-Validation Loop}
    \begin{itemize}
        \item Initialize lists to store the fold-specific estimates of \(\hat{\psi}_1\) and \(\hat{\psi}_0\).
        \item For each fold \(j\) (from 1 to \(k\)):
        \begin{itemize}
            \item \textbf{Training Set:} Combine all folds except \(D_j\) to create the training set \( \{(X_i, A_i, Y_i)\}_{i \in \text{Training Set}} \).
            \item \textbf{Validation Set:} Use fold \(D_j\) as the validation set \( \{(X_i, A_i, Y_i)\}_{i \in \text{Validation Set}} \).
        \end{itemize}
    \end{itemize}

    \item \textbf{Estimate Propensity Scores in the Training Set}
    \begin{itemize}
        \item Fit a propensity score model \(\hat{\pi}(X)\) using the training set.
        \item Calculate the estimated propensity scores \(\hat{\pi}(X_i)\) for all \(i\) in the validation set.
    \end{itemize}

    \item \textbf{Estimate Outcome Regressions in the Training Set}
    \begin{itemize}
        \item Fit outcome regression models \(\hat{\mu}_0(X)\) and \(\hat{\mu}_1(X)\) using the training set.
        \item Calculate the predicted outcomes \(\hat{\mu}_0(X_i)\) and \(\hat{\mu}_1(X_i)\) for all \(i\) in the validation set.
    \end{itemize}

    \item \textbf{Calculate the Doubly Robust Estimator in the Validation Set}
    \begin{itemize}
        \item Initialize two variables to accumulate the contributions from the treated and control groups in the validation set: \( \hat{\psi}_1^j \) and \( \hat{\psi}_0^j \).
        \item For each observation \(i\) in the validation set:
        \begin{itemize}
            \item Compute the contribution for the treated group:
            \[
            \hat{\psi}_{1i} = \frac{A_i}{\hat{\pi}(X_i)} \left( Y_i - \hat{\mu}_1(X_i) \right) + \hat{\mu}_1(X_i)
            \]
            \item Compute the contribution for the control group:
            \[
            \hat{\psi}_{0i} = \frac{1 - A_i}{1 - \hat{\pi}(X_i)} \left( Y_i - \hat{\mu}_0(X_i) \right) + \hat{\mu}_0(X_i)
            \]
            \item Accumulate the contributions:
            \[
            \hat{\psi}_1^j \leftarrow \hat{\psi}_1^j + \hat{\psi}_{1i}
            \]
            \[
            \hat{\psi}_0^j \leftarrow \hat{\psi}_0^j + \hat{\psi}_{0i}
            \]
        \end{itemize}
        \item Calculate the averages for the validation set:
        \[
        \hat{\psi}_1^j = \frac{1}{n_j} \sum_{i \in \text{Validation Set}} \hat{\psi}_{1i}
        \]
        \[
        \hat{\psi}_0^j = \frac{1}{n_j} \sum_{i \in \text{Validation Set}} \hat{\psi}_{0i}
        \]
        \item Store the fold-specific estimates. 
    \end{itemize}

    \item \textbf{Aggregate Results Across Folds}
    \begin{itemize}
        \item Calculate the overall estimates by averaging the fold-specific estimates:
        \[
        \hat{\psi}_1 = \frac{1}{k} \sum_{j=1}^k \hat{\psi}_1^j
        \]
        \[
        \hat{\psi}_0 = \frac{1}{k} \sum_{j=1}^k \hat{\psi}_0^j
        \]
    \end{itemize}

    \item \textbf{Compute the Average Treatment Effect (ATE)}
    \begin{itemize}
        \item Estimate the ATE:
        \[
        \hat{\psi} = \hat{\psi}_1 - \hat{\psi}_0
        \]
    \end{itemize}

    \item \textbf{Output}
    \begin{itemize}
        \item The doubly robust estimator of the average treatment effect \( \hat{\psi} \).
    \end{itemize}

\end{enumerate}
\subsection{Further Discussions}
So far, we have introduced methods yielding efficient influence functions and estimators and given the detailed algebraic transformation for the efficient/doubly robust causal estimator. However, in some cases, the underlying estimand is not that precise, and the decomposition of its tangent space is not feasible; also, in some cases, we only have a very small sample size, and the information is limited. If so, we can only use the empirical cases to derive the heuristic approximation for its efficient estimator (which is close but not attained, the Cramer-Rao bound). We will give an example in the next chapter when we need to derive the efficient/doubly robust estimation for the left-truncated-right-censored survival data.
\section{Conclusion}
In this chapter, we introduce the basic ideas and mathematical tools to perform the causal inference in an efficient/doubly robust way from the observational/ survey data in social science. Intrinsically, to correctly identify the causal estimand from the observational data (via the statistical estimand), for all the methods, no matter whether it is a parametric (regressive) model, a nonparametric (machine learning) model, or a semi-parametric (bias reduction) model, the key is the same: to correctly specify/identify the two models: the propensity score model which allocates cases into the treatment and control group and the outcome result model which specifies the factual or counterfactual outcomes. \\

The doubly robust estimator provides a toolbox that, compared to the previous models, requires the correct specification of both the propensity score model and the outcome result model; we only need to correctly specify one of them. This makes social science research, especially under theoretical-driven studies, much more convenient to yield an unbiased and efficient estimator for the treatment effect. In empirical studies, researchers are always afraid of omitted variable bias when specifying the outcome model, but if the researchers' theories and hypothesis could make the propensities allocating to the treatment and control groups deterministic, then the results are robust (this is pretty like the idea behind the local average treatment effect estimation). \\

In the chapters below, we will again use the asymptotical analysis methods (with the score and influence function) to yield efficient estimators under different data structures and model settings. In this chapter, we just give readers from social science backgrounds a preliminary introduction (with necessary mathematical transforms) to this area. The method for semiparametric doubly robust target double machine learning \citep{kennedy2022review} is definitely one of the fastest developing areas in statistics, econometrics, data science, and relevant discipline's methodological discussions. Like other disciplines, social scientists and demographers need the toolbox to have better causal estimations of their research interests. 
\bibliographystyle{chicago}
\bibliography{try.bib}
\end{document}